\documentclass[12pt, letterpaper]{article}
\pdfoutput=1
\usepackage{amsmath,amssymb,amsfonts,cite,color,epsf,epsfig,epstopdf,graphics,graphicx,mathtools,subcaption,physics,verbatim}

\usepackage[utf8]{inputenc}

\usepackage{ytableau}
\ytableausetup
{mathmode, smalltableaux, aligntableaux = center}

\def\thefootnote{*\arabic{footnote}}
\definecolor{ultramarine}{rgb}{0.07, 0.04, 0.56}
\definecolor{cadmiumgreen}{rgb}{0.0, 0.42, 0.24}
\definecolor{indigo(dye)}{rgb}{0.0, 0.25, 0.42}
\usepackage[linktocpage=true,breaklinks]{hyperref}
\hypersetup{
	colorlinks=true,
	citecolor=ultramarine,
	linkcolor=cadmiumgreen,
	urlcolor=indigo(dye),
}

\numberwithin{equation}{section}

\usepackage{array}
\newcolumntype{P}[1]{>{\centering\arraybackslash}p{#1}}

\usepackage{dcolumn}
\newcolumntype{M}[1]{>{\centering\arraybackslash}m{#1}}
\newcolumntype{N}{@{}m{0pt}@{}}

\newcommand{\g}{g_{M}}
\newcommand{\Mpl}{M_*}
\newcommand{\gf}{g_{\rm eff}}
\newcommand{\eff}{{\rm eff}}
\newcommand{\cutoff}{\Lambda_*}

\newcommand{\spatialR}{ {}^{(3)}\!\tilde{R} }
\newcommand{\spatialRST}{ {}^{(3)}\!R }
\newcommand{\barL}{\bar{\mathcal{L}}}

\newcommand{\nn}{\nonumber \\}

\newcommand{\D}{{\rm d}}
\newcommand{\brp}[1]{\left( #1 \right)}  
\newcommand{\brb}[1]{\left[ #1 \right]}  
\newcommand{\brc}[1]{\left\{ #1 \right\}}  
\newcommand{\be}{\begin{equation}}  
\newcommand{\ee}{\end{equation}}

\usepackage{tikz}

\begin{document}

\begin{flushright} {\footnotesize YITP-21-132\\
IPMU21-0079}  \end{flushright}
\vspace{0.5cm}

\begin{center}

\def\thefootnote{\fnsymbol{footnote}}

{\Large {\bf The Effective Field Theory of Vector-Tensor Theories}}
\\[1cm]

{Katsuki Aoki$^{1}$, Mohammad Ali Gorji$^{1}$,
Shinji Mukohyama$^{1,2}$, Kazufumi Takahashi$^{1}$}
\\[.7cm]

{\small \textit{$^1$Center for Gravitational Physics, Yukawa Institute for Theoretical
		Physics, Kyoto University, 606-8502, Kyoto, Japan
}}\\

{\small \textit{$^2$Kavli Institute for the Physics and Mathematics of the Universe (WPI), The University of Tokyo, 277-8583, Chiba, Japan}}

\end{center}

\vspace{.8cm}

\hrule \vspace{0.3cm}

\begin{abstract} 
We investigate a systematic formulation of vector-tensor theories based on the effective field theory (EFT) approach. The input of our EFT is that the spacetime symmetry is spontaneously broken by the existence of a preferred \emph{timelike} direction in accordance with the cosmological principle. After clarifying the difference of the symmetry breaking pattern from the conventional EFT of inflation/dark energy, we find an EFT description of vector-tensor theories around the cosmological background. This approach not only serves as a unified description of vector-tensor theories but also highlights universal differences between the scalar-tensor theories and the vector-tensor theories. The theories having different symmetry breaking patterns are distinguished by a phenomenological function and consistency relations between the EFT coefficients. We study the linear cosmological perturbations within our EFT framework and discuss the characteristic properties of the vector-tensor theories in the context of dark energy. In particular, we compute the effective gravitational coupling and the slip parameter for the matter density contrast in terms of the EFT coefficients.
\end{abstract}
\vspace{0.5cm} 

\hrule
\def\thefootnote{\arabic{footnote}}
\setcounter{footnote}{0}

\thispagestyle{empty}

\newpage
\hrule
\tableofcontents
\vspace{0.7cm}
\hrule

\newpage
\section{Introduction and summary}\label{introduction}
According to the cosmological observations, the Universe is expanding in time and is spatially homogeneous and isotropic on large scales. Small fluctuations on top of the homogeneous and isotropic background characterize the observed structures in the Universe \cite{Planck:2018vyg}. In the standard big bang cosmology, the Universe keeps to respect these minimal requirements while evolving from the early inflationary era to the late-time dark energy dominated era. However, there are many cosmological models for the different epochs of the Universe. The effective field theory (EFT) approach provides a unified description of theories having the same symmetry breaking pattern which is extremely useful to extract model-independent predictions. In the EFT framework, cosmological models are distinguished by the way how the spacetime symmetry is broken in accordance with the cosmological observations.

The EFT for cosmology was first developed in the context of inflation~\cite{Creminelli:2006xe,Cheung:2007st} by generalizing the EFT of ghost condensation~\cite{Arkani-Hamed:2003pdi,Arkani-Hamed:2003juy} to the Friedmann-Lema\^itre-Robertson-Walker~(FLRW) backgrounds and later applied to the late-time cosmic acceleration~\cite{Creminelli:2008wc,Gubitosi:2012hu,Bloomfield:2012ff,Gleyzes:2013ooa,Gleyzes:2014rba,Frusciante:2015maa,Frusciante:2016xoj,Lagos:2016wyv,Lagos:2017hdr} (see \cite{Frusciante:2019xia} for a review). The conventional EFT of inflation/dark energy assumes that the spacetime symmetry is spontaneously broken by a preferred spacelike \emph{slicing}. The slicing is specified by a function~$\tilde{t}(x)$ with timelike gradient which may be interpreted as a condensate of a time-dependent scalar field. In the unitary gauge where the time coordinate is identified with $\tilde{t}$, the spacetime diffeomorphism invariance is downgraded to the spatial diffeomorphism invariance. The EFT action is systematically constructed under the residual symmetries of the unbroken spatial diffeomorphisms~(diffs). On the other hand, there would be no a priori reason to only choose the symmetry breaking pattern characterized by a single scalar field. The purpose of this paper is to examine another symmetry breaking pattern and establish a systematic formulation of the corresponding EFT.

\begin{figure}[ht]
\begin{center}
\begin{tikzpicture}
	\node[rectangle,thick,draw=black,text width=6.5cm,text centered,minimum height=1cm] at (0,0){\large{EFT of vector-tensor theories}\\ \large{\bf{[present work]}}};
	\node[rectangle,thick,draw=black,text width=6.5cm,text centered,minimum height=1cm] at (0,3){\large{Shift-symmetric}\\ \large{scalar-tensor theories}\large{\cite{Finelli:2018upr}}};
	\node[rectangle,thick,draw=black,text width=6.5cm,text centered,minimum height=1cm] at (0,6){\large{EFT of inflation/dark energy}\\ \large{\cite{Creminelli:2006xe,Cheung:2007st,Creminelli:2008wc,Gubitosi:2012hu,Bloomfield:2012ff,Gleyzes:2013ooa,Gleyzes:2014rba,Frusciante:2015maa,Frusciante:2016xoj,Lagos:2016wyv,Lagos:2017hdr,Frusciante:2019xia}}};
	\node[rectangle,thick,draw=black,text width=5cm,text centered,minimum height=1cm] at (11,0){\large{Gauged ghost}\\ \large{condensation~\cite{Cheng:2006us,Mukohyama:2006mm}}};
	\node[rectangle,thick,draw=black,text width=5cm,text centered,minimum height=1cm] at (11,6){\large{Ghost condensation}\\ \large{\cite{Arkani-Hamed:2003pdi,Arkani-Hamed:2003juy}}};
	\draw[->,>=stealth,thick,dashed] (3.5,-0.2)--(8.25,-0.2);
	\node[rectangle,text width=5cm,text centered,minimum height=1cm] at (5.875,-0.5){Minkowski};
	\draw[->,>=stealth,thick] (8.25,0.2)--(3.5,0.2);
	\node[rectangle,text width=5cm,text centered,minimum height=1cm] at (5.875,0.5){extension to FLRW};
	\draw[->,>=stealth,thick,dashed] (3.5,5.8)--(8.25,5.8);
	\node[rectangle,text width=5cm,text centered,minimum height=1cm] at (5.875,5.5){Minkowski or de Sitter};
	\draw[->,>=stealth,thick] (8.25,6.2)--(3.5,6.2);
	\node[rectangle,text width=5cm,text centered,minimum height=1cm] at (5.875,6.5){extension to FLRW};
	\draw[->,>=stealth,thick,dashed] (-0.2,0.7)--(-0.2,2.35);
	\node[rectangle,text width=2cm,minimum height=1cm,align=right] at (-1.5,1.525){weak\\coupling\\limit};
	\draw[->,>=stealth,thick] (0.2,2.35)--(0.2,0.7);
	\node[rectangle,text width=3cm,minimum height=1cm,align=left] at (2,1.525){gauging\\shift symmetry};
	\draw[->,>=stealth,thick,dashed] (-0.2,3.65)--(-0.2,5.3);
	\node[rectangle,text width=3cm,minimum height=1cm,align=right] at (-2,4.525){soft-breaking\\shift symmetry};
	\draw[->,>=stealth,thick] (0.2,5.3)--(0.2,3.65);
	\node[rectangle,text width=3cm,minimum height=1cm,align=left] at (2,4.525){imposing\\shift symmetry};
	\draw[->,>=stealth,thick,dashed] (10.1,0.7)--(10.1,5.3);
	\node[rectangle,text width=2cm,minimum height=1cm,align=right] at (8.8,3){weak\\coupling\\limit};
	\draw[->,>=stealth,thick] (10.5,5.3)--(10.5,0.7);
	\node[rectangle,text width=3cm,minimum height=1cm,align=left] at (12.3,3){gauging\\shift symmetry};
\end{tikzpicture}
\caption{The web of EFTs.}
\label{fig:webofeft}
\end{center}
\end{figure}
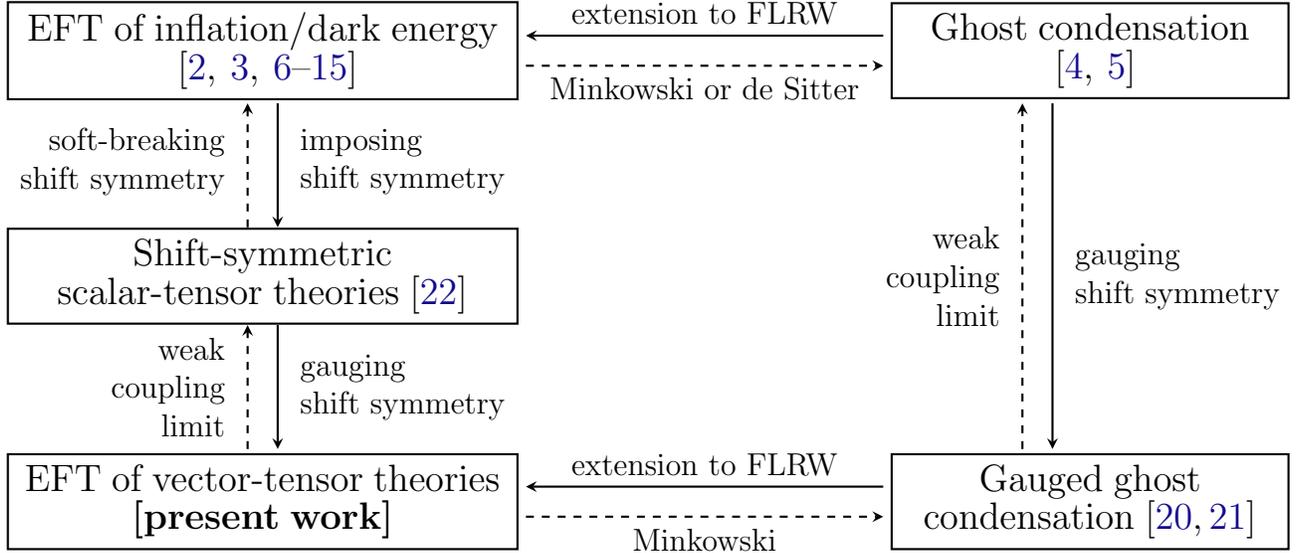

We assume the existence of a preferred \emph{direction} given by a \emph{timelike} vector~$v^{\mu}(x)$.\footnote{An EFT with a timelike vector was studied in \cite{Lagos:2016wyv,Lagos:2017hdr} but it is applicable only for linear perturbations. On the other hand, the EFT formulated in the present paper is systematic, universal, valid at the fully non-linear level and therefore is applicable to perturbations up to any order. EFTs with a spacelike vector were also studied in~\cite{Abolhasani:2015cve,Rostami:2017wiy,Gong:2019hwj} in the context of the so-called anisotropic inflation~\cite{Watanabe:2009ct}.} As is well known, there is no hypersurface orthogonal to a vector in general, so the existence of a preferred direction is inequivalent to the existence of a preferred slicing. The difference becomes transparent by writing the vector as $v_{\mu}=\partial_{\mu}\tilde{t}+\g A_{\mu}$ where $\tilde{t}$ is the St\"{u}ckelberg field associated with the $U(1)$ transformation of the new vector field~$A_{\mu}$ and $\g$ is the gauge coupling. The St\"{u}ckelberg field~$\tilde{t}$ can be chosen to coincide with the time coordinate, and we shall call this choice of the time coordinate the unitary gauge. Thanks to the gauge field~$A_{\mu}$, there exists an additional local symmetry on top of the unbroken spatial diffs in the unitary gauge, yielding a different symmetry breaking pattern from the conventional EFT of inflation/dark energy. This symmetry breaking pattern was studied in gauged ghost condensation~\cite{Cheng:2006us,Mukohyama:2006mm} in the maximally symmetric spacetimes by localizing (gauging) the global shift symmetry of the ghost condensation~\cite{Arkani-Hamed:2003pdi,Arkani-Hamed:2003juy}. The present paper only uses the spatial homogeneity and isotropy as the symmetry of the background spacetime and the EFT will be formulated in generic cosmological backgrounds. Figure~\ref{fig:webofeft} summarizes the relations between different EFTs. 

The preferred vector can be viewed as a condensate of a vector field (but not necessary to be). Various theories with a vector field have been proposed, namely gauged ghost condensation~\cite{Cheng:2006us,Mukohyama:2006mm}, generalized Proca~\cite{Heisenberg:2014rta,Allys:2015sht,BeltranJimenez:2016rff,Allys:2016jaq}, beyond generalized Proca~\cite{Heisenberg:2016eld}, Proca-Nuevo~\cite{deRham:2020yet}, extended vector-tensor~\cite{Kimura:2016rzw}, framids~\cite{Nicolis:2015sra}, and Einstein-aether~\cite{Jacobson:2008aj}. Our EFT approach may serve as a unified description of these vector-tensor theories in the symmetry breaking background that we have mentioned. Moreover, it includes the EFT of scalar-tensor theories and provides a clear comparison with the conventional EFT of inflation/dark energy. 

For the convenience of the reader, we summarize below the essential ingredients of our vector-tensor EFT and the main results:
\begin{itemize}
    \item \emph{Symmetry breaking pattern.} We assume that the spacetime symmetry is spontaneously broken by a timelike preferred vector denoted by $\tilde{\delta}^0{}_{\mu}=\delta^0{}_{\mu}+\g A_{\mu}$ in the unitary gauge~$\tilde{t}=t$. The residual symmetries are the spatial diffs, $\vec{x} \to \vec{x}'(t,\vec{x})$, and the combined $U(1)$ and time diffs, 
    \begin{align}
    t \to t -\g\chi(x)\,, \quad A_{\mu} \to A_{\mu}+\partial_{\mu} \chi(x)
    \,.
    \label{U(1)time_intro}
    \end{align}
    
    \item \emph{Building blocks.} Let $\tilde{g}^{00}=\tilde{\delta}^0{}_{\mu}\tilde{\delta}^0{}_{\nu}g^{\mu\nu}$ and $\tilde{n}^{\mu}=\tilde{\delta}^{\mu}_0/\sqrt{-\tilde{g}^{00}}$ be the norm of the preferred vector and the unit vector along the preferred direction, respectively, where $g_{\mu\nu}$ is the four-dimensional metric and $g^{\mu\nu}$ is its inverse. The projection tensor is denoted by ${\tilde h}_{\mu\nu} = g_{\mu\nu} + {\tilde n}_{\mu} {\tilde n}_{\nu}$. The first derivative of $\tilde{n}^{\mu}$ is decomposed into kinematical quantities, namely the expansion tensor~$\tilde{K}_{\mu\nu}$, the vorticity tensor~$\tilde{\omega}_{\mu\nu}$, and the acceleration vector~$\tilde{a}_{\mu}$. Although $\tilde{n}^{\mu}$ is not a vector normal to a spacelike hypersurface, we can define an object analogous to the spatial curvature, denoted by $\spatialR_{\mu\nu\rho\sigma}$, which we call the orthogonal spatial curvature. The expansion scalar and the orthogonal spatial Ricci scalar are defined by $\tilde{K}=\tilde{K}_{\mu\nu}g^{\mu\nu}=\tilde{K}_{\mu\nu}\tilde{h}^{\mu\nu}$ and $\spatialR=\spatialR_{\mu\nu\rho\sigma}g^{\mu\rho}g^{\nu\sigma}=\spatialR_{\mu\nu\rho\sigma}\tilde{h}^{\mu\rho}\tilde{h}^{\nu\sigma}$, respectively. Essentially, the metric, the norm of the preferred vector, the normalized preferred vector, the kinematical quantities, the orthogonal spatial curvature, and their derivatives are the building blocks of the EFT, but it is convenient to use the field strength~$F_{\mu\nu}=2\partial_{[\mu}A_{\nu]}$ as a part of the building blocks. The field strength can be split into the magnetic part~$\tilde{F}_{\mu\nu}$ and the electric part~$\tilde{F}_{\mu}$ by taking the projections, which are related to the vorticity, the acceleration, and the norm of the preferred vector. See subsection~\ref{sec:building_blocks} for the details.
    
    \item \emph{The EFT action.} We find that the unitary gauge action under the present symmetry breaking pattern is given by 
    \begin{align}
    S&=\int \D^4x \sqrt{-g}\mathcal{L}_{\rm DE} + S_{\rm m}[\psi,g]
    \,, \nn
    \mathcal{L}_{\rm DE}&= \frac{\Mpl^2}{2}f(t) \tilde{R} - \Lambda(t)-c(t) \tilde{g}^{00}-d(t)\tilde{K} + \mathcal{L}_{\rm DE}^{(2)}
    \,,\nn
    \mathcal{L}_{\rm DE}^{(2)}&=
    \frac{1}{2} M_2^4(t) \left(\frac{\delta\tilde{g}^{00}}{-\tilde{g}_{\rm BG}^{00} }\right)^2
    - \frac{1}{2} {\bar M}_1^3(t) \left(\frac{\delta\tilde{g}^{00}}{-\tilde{g}_{\rm BG}^{00} }\right) \delta {\tilde K}
    - \frac{1}{2} {\bar M}_2^2(t) \delta {\tilde K}^2
    - \frac{1}{2} {\bar M}_3^2(t) \delta \tilde{K}^{\alpha}{}_{\beta} \delta \tilde{K}^{\beta}{}_{\alpha} 
    \nonumber \\ 
    & \quad + \frac{1}{2}\mu_1^2(t)\left(\frac{\delta\tilde{g}^{00}}{-\tilde{g}_{\rm BG}^{00} }\right)\delta \spatialR 
    - \frac{1}{4} \gamma_1(t) F_{\alpha\beta} F^{\alpha\beta}  
    - \frac{1}{4} \gamma_2(t) {\tilde F}_{\alpha\beta} {\tilde F}^{\alpha\beta} 
    \nn
    & \quad +\frac{1}{2}\bar{M}_{4}(t) \delta \tilde{K}  \delta \spatialR
    +\frac{1}{2}\lambda_{1}(t) \delta \spatialR^2
    + \cdots \,,
    \label{EFT_intro}
    \end{align}
    where the ellipsis stands for terms which are higher order in either perturbations or derivatives and $S_{\rm m}$ is the matter action, in which the matter fields are collectively denoted by $\psi$. We assume that the matter fields are minimally coupled to the spacetime metric, i.e.~the weak equivalence principle is satisfied. Here, $\tilde{R}=R+2\nabla_{\mu}(\tilde{a}^{\mu}-\tilde{K}\tilde{n}^{\mu})$ is the four-dimensional Ricci scalar with the divergence term subtracted. The subscript BG denotes the background values of corresponding quantities and the quantities~$\delta\tilde{g}^{00}, \delta \tilde{K}_{\mu\nu}$, and $\delta \spatialR$ are the perturbations around their background values. Although we only focus on the lower-order terms of the EFT action, the action can be systematically extended into higher orders. See subsection~\ref{subsec-EFT-action} and also subsections~\ref{subsec-Claas-BB} and \ref{sec:expansion} for the details.

    \item \emph{Consistency relations.} The EFT coefficients in \eqref{EFT_intro} are \emph{not} arbitrary functions of the time~$t$ due to the enlarged symmetry of the EFT. We will show that the necessary and sufficient condition for the invariance under \eqref{U(1)time_intro} leads to consistency relations between the EFT coefficients. Assuming a power-counting, as we will explain in subsections~\ref{sec:expansion} and \ref{subsec-EFT-action}, the consistency relations read
    \begin{align}
    \dot{\Lambda}+3H(\dot{d}+\Mpl^2\dot{f} H) - \frac{1}{2}\Mpl^2 \dot{f} \spatialR_{\rm BG} + \dot{c} \tilde{g}^{00}_{\rm BG} &\simeq 0
    \,,  \label{consistency1_intro} \\
    2M_2^4 \frac{\D}{\bar{N} \D t} \ln (-\tilde{g}^{00}_{\rm BG}) + 3\bar{M}_1^3 \dot{H} - \mu_1^2 \frac{\D}{\bar{N} \D t} \spatialR_{\rm BG} + 2 \dot{c} \tilde{g}^{00}_{\rm BG}
    &\simeq 0
    \,, \\
    \dot{d}+2\Mpl^2 \dot{f} H-3\dot{H}\left(\bar{M}_2^2 + \frac{1}{3}\bar{M}_3^2\right) + \frac{1}{2}\bar{M}_1^3 \frac{\D}{\bar{N} \D t} \ln (-\tilde{g}^{00}_{\rm BG})
    &\simeq 0
    \,,  \\
    \Mpl^2 \dot{f} + \mu_1^2 \frac{\D}{\bar{N} \D t} \ln (-\tilde{g}^{00}_{\rm BG}) &\simeq 0
    \,, \label{consistency4_intro}
\end{align}
where a dot denotes derivative with respect to the cosmic time, e.g.~$\dot{\Lambda}=\D \Lambda/(\bar{N}\D t)$, and $A \simeq B$ means that $A-B$ is suppressed by the cutoff scale of the EFT. Here, $\bar{N}$ and $H$ are respectively the lapse function and the Hubble expansion rate of the background. 

    \item \emph{Comparison with the conventional EFT of inflation/dark energy.} Even though the action~\eqref{EFT_intro} resembles the conventional EFT action, there are universal distinctions between them. The difference is twofold. First, we emphasize that $\tilde{g}^{00}, \tilde{K}_{\mu\nu}$, and $ \spatialR_{\mu\nu\rho\sigma}$ are not the time-time component of the metric, the extrinsic curvature, and the spatial curvature of the constant-time slice, respectively. They are defined with respect to the preferred vector~$\tilde{\delta}^{0}{}_{\mu}=\delta^0{}_{\mu}+\g A_{\mu}$, not the preferred slice specified by $\delta^0{}_{\mu}=\partial_{\mu}t$. Clearly, the parameter~$\g$ characterizes the deviation and we will find that phenomenological differences between scalar-tensor theories and vector-tensor theories are captured by a time-dependent function associated with $\g$ which we will denote by $\alpha_V(t)$. Secondly, as we mentioned, the consistency relations~\eqref{consistency1_intro}--\eqref{consistency4_intro} are required by the residual symmetry in the vector-tensor EFT. However, the EFT coefficients are independent functions in the conventional EFT of inflation/dark energy. Based on these two criteria, we classify the different theories to the three general categories: (i) The EFT of vector-tensor theories: $\alpha_V(t)\neq0$ and the consistency relations should hold; (ii) The EFT of shift-symmetric scalar-tensor theories: $\alpha_V(t)=0$ and  the consistency relations should hold; (iii) The conventional EFT of inflation/dark energy: $\alpha_V(t)=0$ and the consistency relations do not need to hold. See also Figure~\ref{fig:webofeft}.
\end{itemize}

The rest of the paper is organized as follows. In Sec.~\ref{sec:formulation}, we provide a systematic formulation of the EFT of vector-tensor theories. We will elaborate on the symmetry breaking pattern, the building blocks, the EFT action, and the consistency relations. In Sec.~\ref{sec:dictionary}, as a simple demonstration, we discuss how a subclass of the generalized Proca theory is translated into our EFT language. We then give detailed comparisons between scalar-tensor theories and vector-tensor theories by using our EFT formulation in Sec.~\ref{sec:comparison}. In particular, in Sec.~\ref{sec:pheno_prediction}, we formulate a unified description of linear scalar (and tensor) perturbations by means of the so-called $\alpha$-basis~\cite{Bellini:2014fua,Gleyzes:2014qga,Gleyzes:2014rba,Frusciante:2016xoj}. The vector perturbations are separately discussed in Sec.~\ref{sec:vector_pert}. We conclude in Sec.~\ref{summary}. In Appendix~\ref{app:1+3}, we summarize technical details of the $1+3$ decomposition with respect to a general vector with vorticity. Finally, we briefly discuss possible new operators of the vector-tensor EFT arising from Lie derivatives of the building blocks in Appendix~\ref{sec:Lie}.

\section{Formulation}
\label{sec:formulation}

\subsection{Symmetry breaking pattern}
\label{sec:symmetry}

The input of the conventional EFT of inflation/dark energy, which we call the scalar-tensor EFT to distinguish it from the vector-tensor EFT, is the existence of a preferred slicing of spacetime determined by a function~$\tilde{t}(x)$ with timelike gradient, which breaks a part of the spacetime symmetries. In the present paper, we instead assume the existence of a preferred vector~$v_{\mu}(x)$ which is supposed to be non-vanishing and timelike. The preferred vector spontaneously breaks a part of the spacetime symmetries. 

It would be convenient to use the St\"{u}ckelberg trick in order to make a comparison between the scalar-tensor EFT and the present vector-tensor EFT. Let us write the preferred vector as
\begin{align} \label{preferred_v}
    v_{\mu}=\partial_{\mu} \tilde{t} + \g A_{\mu}; \quad \g \equiv \frac{g}{\Mpl} \,,
\end{align}
where $\Mpl$ is the ``Planck mass''\footnote{We use $\Mpl$ to denote the Planck mass rather than the usual notation~$M_{\rm Pl}$ because the observed Planck mass will be different from $\Mpl$.}  and we shall call $g$ (or $\g$) the gauge coupling. The preferred vector is invariant under the residual $U(1)$ transformation,
\begin{align} \label{U(1)_Stu}
\tilde{t} \to \tilde{t} - \g \chi\,, \quad A_{\mu} \to A_{\mu}+\partial_{\mu}\chi
\,,
\end{align}
with an arbitrary scalar~$\chi(x)$. We then define a slicing of the spacetime by using the St\"{u}ckelberg field $\tilde{t}$. The time coordinate~$t$ is chosen to coincide with $\tilde{t}$ so that
\begin{eqnarray}\label{delta-tilde}
v_{\mu} = {\tilde \delta}^0{}_{\mu} \equiv { \delta}^0{}_{\mu} + \g A_{\mu} \,.
\end{eqnarray}
We use this time coordinate throughout the present paper. It is now transparent that the preferred vector does not coincide with the time direction and it is not hypersurface orthogonal in general. Therefore, we deal with a symmetry breaking pattern which is different from the one in the scalar-tensor EFT.

We now clarify the residual symmetries of the vector-tensor EFT. Ones of the residual symmetries are the spatial diffs. In addition, we have the $U(1)$ invariance due to the St\"{u}ckelberg trick. Since the St\"{u}ckelberg field is identified with the time coordinate, the $U(1)$ transformation~\eqref{U(1)_Stu} also yields a change of the time coordinate. Therefore, the residual symmetries are the spatial diffs and a combination of the $U(1)$ and time diffs \eqref{U(1)time_intro}, which are exactly the same as the symmetries of the gauged ghost condensate~\cite{Cheng:2006us,Mukohyama:2006mm}.

Following the gauged ghost condensate, the preferred vector~\eqref{preferred_v} can be identified with the covariant derivative of the field~$\tilde{t}(x)$ where the shift symmetry of $\tilde{t}(x)$ is gauged by the compensating field~$A_{\mu}(x)$. From this perspective, the vector-tensor EFT can be obtained by gauging the scalar-tensor EFT with the global shift symmetry. (See Figure~\ref{fig:webofeft} for the relation between different EFTs.)

In summary, although the vector-tensor EFT we shall consider in the present paper has similarities to the shift-symmetric scalar-tensor EFT, there are clear distinctions. From the geometric point of view, the difference is whether the spacetime symmetry is broken by a preferred slicing or a preferred vector. The difference can be also understood by whether the shift symmetry is global or local.

\subsection{EFT building blocks}
\label{sec:building_blocks}

The spacetime metric~$g_{\mu\nu}$ and the curvature~$R_{\mu\nu\rho\sigma}$ are obviously building blocks of the EFT. In addition, we introduce the invariant tensors to characterize the properties of the preferred vector.

Let us first define the norm of the preferred vector,
\begin{eqnarray}\label{g00-tilde}
{\tilde g}^{00} \equiv {\tilde \delta}^0{}_{\alpha} {\tilde \delta}^0{}_{\beta} g^{\alpha\beta}
=  g^{00} + 2 \g A^0 + \g^2 A_{\alpha} A^\alpha \,,
\end{eqnarray}
which is supposed to be negative. Then, we find the unit timelike vector,
\begin{eqnarray}\label{n-tilde}
{\tilde n}_\mu \equiv -\frac{{\tilde \delta}^0{}_{\mu}}{\sqrt{-{\tilde g}^{00}}} \,, \quad
{\tilde n}_\alpha {\tilde n}^\alpha = -1 \,.
\end{eqnarray}
As we already mentioned, ${\tilde n}_{\mu}$ is not the vector normal to the surfaces of constant $t$ due to the existence of the transverse degrees of freedom of the gauge field~$A_{\mu}$. Nonetheless, we can define the projection tensor,
\begin{eqnarray}\label{h-tilde}
{\tilde h}_{\mu\nu} \equiv g_{\mu\nu} + {\tilde n}_{\mu} {\tilde n}_{\nu} \,,
\end{eqnarray}
which satisfies ${\tilde h}_{\mu}{}^{\alpha} {\tilde h}_{\alpha}{}^\nu = {\tilde h}_{\mu}{}^\nu$ and ${\tilde h}_{\mu}{}^{\alpha}{\tilde n}_\alpha = 0$. By the use of $\tilde{n}^{\mu}$ and $\tilde{h}_{\mu\nu}$, we can define parallel and orthogonal objects with respect to the preferred vector. We have summarized the $1+3$ decomposition with respect to the general unit vector~$\tilde{n}^{\mu}$ in Appendix~\ref{app:1+3}.

The first derivative of the preferred vector~$\nabla_{\mu}\tilde{\delta}^0{}_{\nu}$ is decomposed into two invariant blocks,
\begin{align}
    \partial_{\mu} \tilde{g}^{00}\,, \quad \nabla_{\mu}\tilde{n}_{\nu}\,,
\end{align}
via the relation~$\nabla_{\mu}\tilde{\delta}^0{}_{\nu}= \partial_{\mu} \tilde{g}^{00} \tilde{n}_{\nu}/(2\sqrt{-\tilde{g}^{00}}) - \sqrt{-\tilde{g}^{00}}\nabla_{\mu}\tilde{n}_{\nu}$
where $\tilde{n}^{\nu}\nabla_{\mu}\tilde{n}_{\nu}=0$. The second piece is further decomposed into the expansion tensor, the vorticity tensor, and the acceleration vector which are defined as
\begin{align}
{\tilde K}_{\mu\nu} &\equiv {\tilde h}_{(\mu|}{}^\alpha \nabla_{\alpha}{\tilde n}_{|\nu)} \,,  
\label{K-tilde} \\
{\tilde \omega}_{\mu\nu} &\equiv {\tilde h}_{[\mu|}{}^\alpha \nabla_{\alpha}{\tilde n}_{|\nu]}
\,, \\
\tilde{a}_{\mu} & \equiv \tilde{n}^{\alpha}\nabla_{\alpha} \tilde{n}_{\mu}
\,, \label{acceleration}
\end{align}
respectively. The tensors~\eqref{K-tilde}--\eqref{acceleration} specify the geometric properties of the preferred vector. On the other hand, the field strength of the gauge field $A_{\mu}$,
\begin{align}
    F_{\mu\nu}\equiv \partial_{\mu}A_{\nu}-\partial_{\nu}A_{\mu}\,,
\end{align}
can be decomposed into two invariant blocks
\begin{align}
F_{\mu\nu} = -2 {\tilde F}_{[\mu} {\tilde n}_{\nu]} + {\tilde F}_{\mu\nu} \,,
\end{align}
where
\begin{align}
    \tilde{F}_{\mu} \equiv \tilde{n}^{\alpha}F_{\mu\alpha}
    \,, \quad
    \tilde{F}_{\mu\nu}\equiv \tilde{h}^{\alpha}{}_{\mu}\tilde{h}^{\beta}{}_{\nu}F_{\alpha\beta}
    \,,
    \label{def_tildeF}
\end{align}
are respectively the ``electric'' and ``magnetic'' parts of the field strength. Clearly, only two of $F_{\mu\nu}$, ${\tilde F}_{\mu}$, ${\tilde F}_{\mu\nu}$ are independent. Moreover, one can also find the following relations
\begin{align}
\tilde{\omega}_{\mu\nu}
&=-\frac{\g}{2\sqrt{-\tilde{g}^{00}}} \tilde{F}_{\mu\nu}
\,, 
\label{omega_rel}
\\
\tilde{a}_\mu&=\frac{\g}{\sqrt{-\tilde{g}^{00}}}\tilde{F}_\mu-\frac{\tilde{h}^{\alpha}{}_{\mu}\partial_{\alpha}\tilde{g}^{00}}{2\tilde{g}^{00}}\,,
\label{a_rel}
\end{align}
meaning that there are redundant operators. Since we will discuss the limit~$\g\to 0$ to make a comparison with the scalar-tensor EFT later, we shall adopt the field strength and omit the vorticity and the acceleration from the independent building blocks of the EFT action.

It is useful to introduce objects analogous to the spatial covariant derivative and the spatial curvature. As summarized in Appendix~\ref{app:1+3}, although there is no preferred spatial hypersurface, we can define the analogous objects associated with the preferred vector, which we shall call the orthogonal spatial covariant derivative and the orthogonal spatial curvature, respectively. The orthogonal spatial derivatives~$\tilde{D}_{\mu}$ is defined by
\begin{align}
    \tilde{D}_{\mu} \tilde{T}^{\nu \cdots}{}_{\rho \cdots} 
    \equiv \tilde{h}^{\alpha}{}_{\mu}\tilde{h}^{\nu}{}_{\beta}\cdots \tilde{h}^{\gamma}{}_{\rho} \cdots
    \nabla_{\alpha} \tilde{T}^{\beta \cdots}{}_{\gamma \cdots} 
    \,,
\end{align}
for a projected tensor~$\tilde{T}^{\nu \cdots}{}_{\rho \cdots}$ which satisfies $\tilde{T}^{\nu \cdots}{}_{\rho \cdots}\tilde{n}_{\nu}=\tilde{T}^{\nu \cdots}{}_{\rho \cdots}\tilde{n}^{\rho}=0$. 
Note that the derivative~$\tilde{D}_{\mu}$ is defined only in the tangent hyperplane at each point and not on a 3-manifold when the vorticity does not vanish (see e.g.~\cite{ellis2012relativistic,Roy:2014lda}). The orthogonal spatial curvature may be defined through the commutator of the orthogonal spatial covariant derivative~\cite{Roy:2014lda}. In practice, the orthogonal spatial curvature~${}^{(3)}\!\tilde{R}_{\mu\nu\rho\sigma}$ can be computed by the use of the Gauss equation,
\begin{align}
    \tilde{h}^{\alpha}{}_{\mu} \tilde{h}^{\beta}{}_{\nu} \tilde{h}^{\gamma}{}_{\rho} \tilde{h}^{\delta}{}_{\sigma} R_{\alpha\beta\gamma\delta} =
    {}^{(3)}\!\tilde{R}_{\mu\nu\rho\sigma} + 2 \tilde{B}_{[\rho|\mu} \tilde{B}_{|\sigma]\nu}
    \,, \label{eqn:Gaussequation}
\end{align}
where $\tilde{B}_{\mu\nu}=\tilde{K}_{\mu\nu}+\tilde{\omega}_{\mu\nu}$. We also define
\begin{align}
    \spatialR_{\mu\nu} \equiv \spatialR^{\alpha}{}_{\mu\alpha \nu}\,, \quad
    \spatialR \equiv \spatialR^{\mu}{}_{\mu} = \spatialR^{\mu\nu}{}_{\mu\nu}
    \,.
\end{align}
All the components of the four-dimensional curvature can be written in terms of the orthogonal spatial curvature and the kinematical quantities [see \eqref{Codazzi_eq} and \eqref{Ricci_eq}], so we can use either $R_{\mu\nu\rho\sigma}$ or ${}^{(3)}\!\tilde{R}_{\mu\nu\rho\sigma}$ as an independent building block. In particular, the four-dimensional Ricci scalar is written as
\begin{align}
    R&={}^{(3)}\! \tilde{R} + \tilde{B}_{\mu\nu}\tilde{B}^{\nu\mu}-\tilde{K}^2  -2 \nabla_{\mu}(\tilde{a}^{\mu} -\tilde{K}\tilde{n}^{\mu})
\,.
\end{align}
The choice is totally a matter of the EFT formulation. In the present context, the orthogonal spatial curvature~${}^{(3)}\!\tilde{R}_{\mu\nu\rho\sigma}$ would be more useful since $R_{\mu\nu\rho\sigma}$ includes time derivatives, leading to an additional would-be ghostly state when we consider higher-order terms of the curvature. In fact, the Lagrangian of the generalized Proca theory~\cite{Heisenberg:2014rta}, a ghost-free vector-tensor theory, is written in a simpler form by the use of ${}^{(3)}\!\tilde{R}_{\mu\nu\rho\sigma}$ as we see in Sec.~\ref{sec:dictionary}. We thus adopt ${}^{(3)}\!\tilde{R}_{\mu\nu\rho\sigma}$ as the building block of our EFT formulation.

The four-dimensional covariant derivative~$\nabla_{\mu}$ is decomposed into the Lie derivative~$\pounds_{\tilde{n}}$ along $\tilde{n}^{\mu}$ and the orthogonal spatial derivatives~$\tilde{D}_{\mu}$. Our EFT building blocks are
\begin{align}
\tilde{g}^{00}, \quad \spatialR_{\mu\nu\rho\sigma}, \quad \tilde{F}_{\mu\nu}, \quad \tilde{F}_{\mu}, \quad \tilde{K}_{\mu\nu}, \quad \pounds_{\tilde{n}}, \quad \tilde{D}_{\mu}
    \,,
    \label{EFTblocks}
\end{align}
and the metric~$g^{\mu\nu}$. Then, all the tensors in \eqref{EFTblocks} have no components parallel to $\tilde{n}^{\mu}$, so one can use either $g^{\mu\nu}$ or $\tilde{h}^{\mu\nu}$ to contract indices. The most general EFT action of the vector-tensor theories is given by
\begin{align}
    S&=\int \D^4x \sqrt{-g} \mathcal{L}_{\rm DE}+ S_{\rm m}[\psi,g]
    \,,
    \\
    \mathcal{L}_{\rm DE}&= 
    \mathcal{L}_{\rm DE} (\tilde{g}^{00}, \spatialR_{\mu\nu\rho\sigma},  \tilde{F}_{\mu\nu}, \tilde{F}_{\mu}, \tilde{K}_{\mu\nu}, \pounds_{\tilde{n}},  \tilde{D}_{\mu})
    \,,
    \label{EFT_most_general}
\end{align}
where the indices are contracted by $g^{\mu\nu}$ (or equivalently by $\tilde{h}^{\mu\nu}$) and $S_{\rm m}$ is the action for matter fields, collectively denoted by $\psi$, which are minimally coupled to the spacetime metric.

The tensors~$\tilde{K}_{\mu\nu}$ and ${}^{(3)}\!\tilde{R}_{\mu\nu\rho\sigma}$ should be distinguished from the extrinsic curvature of the constant-time hypersurface~$K_{\mu\nu}= h_{(\mu|}{}^\alpha \nabla_{\alpha}n_{|\nu)}$ with $n_{\mu} = -\delta^0{}_\mu/\sqrt{-g^{00}}$ and the intrinsic curvature of the constant-time hypersurface~${}^{(3)}\!R_{\mu\nu\rho\sigma}$ since the preferred vector is not orthogonal to the constant-time hypersurface. Nonetheless, it is worth mentioning that they agree when we consider irrotational solutions, i.e.~those without transverse degrees of freedom. In this case, $\tilde{\omega}_{\mu\nu}=0$ and we can write the vector field components as\footnote{Alternatively, we can work with the covariant irrotational ansatz~$A_\mu=\phi_1\partial_\mu \phi_2$ and use the residual $U(1)$ transformation~\eqref{U(1)_Stu} to make the preferred vector orthogonal to the constant-time hypersurfaces. However, the ansatz~\eqref{irr_sol} takes the form~$A_\mu = A_0 \delta^0{}_\mu$ in the gauge~$A=0$ which is more suitable to discuss the cosmological perturbations later.}
\begin{align}
    A_{\mu}=(A_0, \partial_i A)
    \,.
    \label{irr_sol}
\end{align}
We can set $A=0$ by using the residual gauge freedom of the combined $U(1)$ and time diffs. In this gauge choice, the time coordinate is completely fixed except the freedom of the time reparametrization, $t \to t'(t)$. Then, the vector~$\tilde{n}_{\mu}$ coincides with the vector orthogonal to the constant-time hypersurfaces, $n_{\mu}$. Hence, we have
\begin{align}
    \tilde{n}^{\mu} & = n^{\mu}\,, 
    \label{n_irro} 
    \\
    \tilde{D}_{\mu} &= D_{\mu} \,, 
    \\
    \tilde{K}_{\mu\nu} & = K_{\mu\nu} \,, 
    \label{K_irro} 
    \\
    {}^{(3)}\!\tilde{R}_{\mu\nu\rho\sigma} &= {}^{(3)}\! R_{\mu\nu\rho\sigma}\,,
\end{align}
under the irrotational ansatz~\eqref{irr_sol} and the gauge choice, $A=0$, where $D_{\mu}$ is the spatial covariant derivative on the constant-time hypersurface. These relations are practically useful to compute cosmological perturbations. 

\subsection{Classifications of building blocks}\label{subsec-Claas-BB}

A crucial difference between the vector-tensor EFT and the scalar-tensor EFT is the absence of $t$ in the building blocks. In the scalar-tensor EFT, the background parts of the building blocks can be subtracted without breaking the residual gauge freedom; for instance, $\delta K \equiv K - 3H(t)$ is invariant under the residual gauge freedom of the scalar-tensor EFT, where $H(t)$ is the Hubble expansion rate. On the other hand, the similar quantity~$\delta \tilde{K}\equiv \tilde{K}- 3H(t)$ is not an invariant quantity in the present setup. This fact requires a formulation different from the scalar-tensor EFT. As a next step of our EFT formulation, we shall classify the building blocks into those that contribute to the background dynamics and those that do not.

Let us first discuss the background dynamics of the FLRW universe with the preferred vector. Thanks to the homogeneity and the isotropy, at any points, the FLRW spacetime and the background preferred vector are characterized by the spatial curvature, the Hubble expansion rate, the norm of the preferred vector, and their time derivatives. All these quantities can be read by the background values of the building blocks
\begin{align}
    \tilde{g}^{00},\quad \tilde{K},\quad \spatialR,\quad   \pounds_{\tilde{n}},
\label{background_blocks}
\end{align}
where $\tilde{K}\equiv \tilde{K}^{\mu}{}_{\mu}$ is the expansion scalar and $\pounds_{\tilde{n}}$ is the Lie derivative with respect to $\tilde{n}^{\mu}$. The quantities in \eqref{background_blocks} are the essential building blocks to determine the background.

We then subtract the background parts from the other building blocks by the use of \eqref{background_blocks}. The building blocks~${\tilde F}_{\mu\nu}, {\tilde F}_{\mu}$, and ${\tilde D}_{\mu}$ have no background part and there is no need for the subtraction. The subtraction is needed for the expansion tensor, yielding the shear tensor,
\begin{align}
    \tilde{\sigma}_{\mu\nu} \equiv \tilde{K}_{\mu\nu}- \frac{1}{3}\tilde{K}\tilde{h}_{\mu\nu}
    \,.
\end{align}
Similarly, the symmetric trace-free part of the curvature, defined by
\begin{align}
    \spatialR^T_{\mu\nu}\equiv \spatialR_{(\mu\nu)} - \frac{1}{3}\tilde{h}_{\mu\nu}\spatialR
    \,,
\end{align}
starts linear in perturbations around the FLRW background. As detailed in Appendix~\ref{app:1+3}, the orthogonal spatial curvature~$\spatialR_{\mu\nu\rho\sigma}$ is decomposed into three irreducible pieces, $\spatialR$, $\spatialR^T_{\mu\nu}$, and $\spatialR^V_{\mu}\equiv \spatialR_{\mu\nu\rho\sigma}\tilde{\epsilon}^{\nu\rho\sigma}$ with $\tilde{\epsilon}^{\nu\rho\sigma}\equiv \tilde{n}_{\mu}\epsilon^{\mu\nu\rho\sigma}$, where the last piece is expressed in terms of $\tilde{B}_{\mu\nu}$ (see \eqref{eqn:Gaussequation}). Since $\tilde{B}_{\mu\nu}=\tilde{K}_{\mu\nu}+\tilde{\omega}_{\mu\nu}$ and $\tilde{\omega}_{\mu\nu}\propto \tilde{F}_{\mu\nu}$, we only need to use  $\spatialR$ and $\spatialR^T_{\mu\nu}$ as independent building blocks of the EFT. As a result, the independent EFT building blocks that start at the linear or higher order in perturbations around the FLRW background are
\begin{align}
     \tilde{F}_{\mu\nu}\,, \quad \tilde{F}_{\mu}\,, \quad  \tilde{\sigma}_{\mu\nu}\,, \quad \spatialR^T_{\mu\nu}\,, \quad \tilde{D}_{\mu}
    \,.
    \label{pert_blocks}
\end{align}

The EFT Lagrangian \eqref{EFT_most_general} can be written as
\begin{align}
    \mathcal{L}_{\rm DE}=\mathcal{L}_0(\tilde{g}^{00},\tilde{K},\spatialR,\pounds_{\tilde{n}})
    +\mathcal{L}_2(\tilde{g}^{00},\tilde{K},\spatialR,\pounds_{\tilde{n}},
    \tilde{F}_{\mu\nu},\tilde{F}_{\mu},  \tilde{\sigma}_{\mu\nu},\spatialR^T_{\mu\nu},\tilde{D}_{\mu})
    \,.
    \label{EFT_general_covariant}
\end{align}
We can assume that $\mathcal{L}_2$ starts at the quadratic or higher order in perturbations around the FLRW background without loss of generality since \eqref{pert_blocks} cannot form a scalar at the linear order by themselves. We can also assume $\mathcal{L}_2$ does not contain the terms of the forms
\begin{align}
    F(\tilde{g}^{00},\tilde{K},\spatialR,\pounds_{\tilde{n}})\pounds_{\tilde{n}}\tilde{T}&=F(\tilde{g}^{00},\tilde{K},\spatialR,\pounds_{\tilde{n}})\tilde{n}^{\mu}\nabla_{\mu}\tilde{T}\,, 
    \label{FLT} \\
    F(\tilde{g}^{00},\tilde{K},\spatialR,\pounds_{\tilde{n}})\tilde{D}_{\mu}\tilde{T}^{\mu}&=F(\tilde{g}^{00},\tilde{K},\spatialR,\pounds_{\tilde{n}})\tilde{h}^{\mu\nu}\nabla_{\mu}\tilde{T}_{\nu}
    \,,
    \label{FDT}
\end{align}
without loss of generality, where $F$ is a function of \eqref{background_blocks}, $\tilde{T}$ is a scalar, and $\tilde{T}^{\mu}$ is an orthogonal vector constructed from the building blocks satisfying $\tilde{T}^{\mu}\tilde{n}_{\mu}=0$. Indeed, performing integration by parts, \eqref{FLT} and \eqref{FDT} become
\begin{align}
    \int \D^4x \sqrt{-g} F \tilde{n}^{\mu}\nabla_{\mu}\tilde{T} &= - \int \D^4x \sqrt{-g} (\pounds_{\tilde{n}}F +\tilde{K} F ) \tilde{T}
    \,, \\
    \int \D^4x \sqrt{-g} F\tilde{h}^{\mu\nu}\nabla_{\mu}\tilde{T}_{\nu}
    &=-\int \D^4x \sqrt{-g} ( \tilde{D}_{\mu}F +\tilde{a}_{\mu} F)\tilde{T}^{\mu}
    \,,
\end{align}
meaning that \eqref{FLT} and \eqref{FDT} can be absorbed into other terms.

So far, we have explicitly preserved the residual symmetries of the EFT. The Lagrangian~\eqref{EFT_general_covariant} is the most general Lagrangian of the vector-tensor EFT. However, \eqref{EFT_general_covariant} is not as useful as that of the scalar-tensor EFT because the Lagrangian is a non-linear function of the building blocks. We need an additional step to bring our general EFT formulation to a more useful form.

\subsection{Expansion around cosmological background}
\label{sec:expansion}
We introduce the quantities
\begin{align} \label{pert_non_scalars}
\delta \tilde{g}^{00}\equiv \tilde{g}^{00}- \tilde{g}^{00}_{\rm BG}(t)\,, \quad
\delta \tilde{K}\equiv \tilde{K}-\tilde{K}_{\rm BG}(t) \,, \quad 
\delta \spatialR \equiv \spatialR - \spatialR_{\rm BG}(t)\,,
\end{align}
where ${\tilde g}^{00}_{\rm BG}(t)$, $\tilde{K}_{\rm BG}(t)$, and $\spatialR_{\rm BG}(t)$ are the background values of the norm of the preferred vector, the expansion scalar, and the orthogonal spatial curvature, respectively. We then perform a Taylor expansion of the Lagrangian to the power of \eqref{pert_non_scalars}. Note that the residual symmetries of the EFT become implicit after the Taylor expansion since the quantities in \eqref{pert_non_scalars} are not invariant under the combined $U(1)$ and time diffs~\eqref{U(1)time_intro}. The symmetry is implemented as consistency relations between the Taylor coefficients as we will see below.

We also perform the derivative expansion of the EFT Lagrangian on top of the perturbative expansion around the FLRW background. The leading-order Lagrangian up to the quadratic order in perturbations is given by
\begin{align}
    \mathcal{L}_0=&\;\barL_0 + \barL_{\tilde{g}^{00}} \delta \tilde{g}^{00} + \barL_{\tilde{K}} \delta \tilde{K} +\barL_{\spatialR} \delta \spatialR
    \nn
    &+ \frac{1}{2}\barL_{\tilde{g}^{00}\tilde{g}^{00}} ( \delta \tilde{g}^{00})^2 + \barL_{\tilde{g}^{00}\tilde{K}}  \delta \tilde{g}^{00} \delta \tilde{K} + \frac{1}{2}\barL_{\tilde{K}\tilde{K}} \delta \tilde{K}^2 
    \nn
    &+\barL_{\tilde{g}^{00}\spatialR}\delta \tilde{g}^{00} \delta \spatialR + \bar{\mathcal{L}}_{\tilde{K}\spatialR} \delta\tilde{K}\delta\spatialR +\frac{1}{2} \bar{\mathcal{L}}_{\spatialR \spatialR} \delta \spatialR ^2 
    +\cdots \,,
    \label{expansionL0}
\end{align}
and
\begin{align}
    \mathcal{L}_2&=\mathcal{L}_{\tilde{F}_{*}^2} \tilde{F}_{\mu}\tilde{F}^{\mu} + \mathcal{L}_{\tilde{F}_{**}^2} \tilde{F}_{\mu\nu}\tilde{F}^{\mu\nu} + \mathcal{L}_{\tilde{\sigma}_{**}^2} \tilde{\sigma}_{\mu\nu}\tilde{\sigma}^{\mu\nu} 
    +\cdots
    \nn
    &=\bar{\mathcal{L}}_{\tilde{F}_{*}^2} \tilde{F}_{\mu}\tilde{F}^{\mu} + \bar{\mathcal{L}}_{\tilde{F}_{**}^2} \tilde{F}_{\mu\nu}\tilde{F}^{\mu\nu} + \bar{\mathcal{L}}_{\tilde{\sigma}_{**}^2} \tilde{\sigma}_{\mu\nu}\tilde{\sigma}^{\mu\nu} 
    +\cdots
    \,,
    \label{expansionL2}
\end{align}
where $\mathcal{L}_0$, $\mathcal{L}_{\tilde{F}_{*}^2}$, $\mathcal{L}_{\tilde{F}_{**}^2}$, and $ \mathcal{L}_{\tilde{\sigma}_{**}^2}$ are functions of \eqref{background_blocks} while the barred quantities are coefficients of the Taylor series. The Taylor coefficients are functions of the background values of \eqref{background_blocks} and then are implicit functions of the time.

Let us show that the operators presented in \eqref{expansionL0} and \eqref{expansionL2} are indeed leading operators of the EFT. We are interested in vector-tensor theories having $1+2+2$ degrees of freedom:~$1$ is the longitudinal mode of the preferred vector, $2$ are the transverse modes of the vector, and the last $2$ are the gravitational tensor degrees of freedom. The kinetic term and the gradient term of the tensor modes are governed by the following operators
\begin{align}
    \tilde{\sigma}_{\mu\nu}\tilde{\sigma}^{\mu\nu}\,, \quad \delta \spatialR \,,
\end{align}
respectively, which are the leading ones. We identify leading operators associated with the scalar and vector degrees of freedom under the decoupling limit of gravity (or the test-field limit of the preferred vector), where the metric fluctuations may be ignored while the Hubble scale of the background metric is kept finite. In the decoupling limit, the metric and the preferred vector are given by
\begin{align}
    g_{\mu\nu}\D x^{\mu}\D x^{\nu}&=-\bar{N}^2(t)\D t^2+a^2(t)\delta_{ij}\D x^i \D x^j\,, \label{decoupling_metric} \\
    \tilde{\delta}^0_{\mu}&=\delta^0_{\mu}+ \g A_{\mu}\,, \label{decoupling_vector}
\end{align}
where, without loss of generality, we can assume $A_{\mu}$ has no background part by using the freedom of the time reparametrization. We then obtain 
\begin{align}
\delta_1 \tilde{g}^{00} &= -\frac{2\g A_0}{\bar{N}^2}
\,, \\
\delta_1 \tilde{K} &= -\g \frac{\bar{N}}{a^2}\partial_i A^i 
\,, \label{eqn:delta1Ktilde} \\
\delta_1\! \spatialR &= 4\g H \frac{\bar{N}}{a^2}\partial_i A^i 
\,, \label{eqn:delta1R3} \\
\delta_1 \tilde{F}_{\mu\nu} &= \begin{pmatrix} 
0 & 0 \\
0 & 2\partial_{[i} A_{j]}
\end{pmatrix}
 \,, \\
\delta_1 \tilde{F}_{\mu}&=
\left(0, \frac{\partial_i A_0}{\bar{N}} - \frac{\D}{\bar{N}\D t} A_i \right) 
\,, \\ 
\delta_1 \tilde{\sigma}_{\mu\nu} &= 
-\g \bar{N} 
\begin{pmatrix}
0 & 0 \\
0 & \partial_{(i} A_{j)} - \frac{1}{3}\delta_{ij}\partial_k A^k
\end{pmatrix}\,,
\end{align}
where $\delta_1 \mathcal{Q}$ denotes the part of $\mathcal{Q}$ linear in $A_{\mu}=(A_0,A_i)$ and $H=\frac{\D a}{\bar{N} \D t}/a$ is the Hubble expansion rate. The EFT Lagrangian~\eqref{expansionL0} contains the operators starting at the linear order in perturbations, i.e.~$\delta \tilde{g}^{00}$, $\delta \tilde{K}$, and $\delta \spatialR$. Whereas the linear parts of $\delta \tilde{K}$ and $\delta \spatialR$ are total divergences, the linear part of $\delta \tilde{g}^{00}$ leads to a tadpole term for $A_0$, meaning that the background equation of motion imposes
\begin{align}
    \barL_{\tilde{g}^{00}}=0
    \,,
\end{align}
provided $\tilde{g}^{00}_{\rm BG} \neq 0$. (See subsection~\ref{sec:background} for the argument showing the same conclusion without relying on the decoupling limit.) To discuss the quadratic Lagrangian, we also need quadratic parts of $\delta \tilde{K}$ and $\delta \spatialR$ which are given by
\begin{align}
    \delta_2 \tilde{K} &=  \g^2 \left[ \frac{1}{2\bar{N}a^3} \partial_t \left(\bar{N}^2 a A_i A^i \right) + \frac{\bar{N}}{a^2} \partial_i (A_0 A^i) \right] 
    \,, \label{eqn:delta2Ktilde}  \\
    \delta_2\! \spatialR &= \g^2 \left[ \frac{-2}{\bar{N}a^3 H} \partial_t \left( H^2 \bar{N}^2 a A_i A^i \right)
    +\frac{\bar{N}^2}{a^4} \partial_i \left( A_j \partial^j A^i -  A^i \partial_j A^j \right)
    -4H \frac{\bar{N}}{a^2}\partial_i (A_0 A^i) \right]
    \,. \label{eqn:delta2R3}
\end{align}
It is convenient to introduce the St\"{u}ckelberg field $\pi$ according to
\begin{align}
    A_0 \to A_0+\g^{-1} \partial_t \pi\,, \quad A_i \to A_i + \g^{-1} \partial_i \pi\,,
\end{align}
and set the gauge $\partial_i A^i=0$.\footnote{We have used $\tilde{t}$ to denote the St\"{u}ckelberg field which has a time-dependent expectation value. Here, the field~$\pi$ has no expectation value and can be regarded as the perturbation of $\tilde{t}$.} Then, the longitudinal mode is represented by $\pi$ while $A_i$ corresponds to the transverse degrees of freedom. The additional variable $A_0$ is non-dynamical. The kinetic terms of $\pi$ and $A_i$ correspond to the operators
\begin{align}
    (\delta \tilde{g}^{00})^2 \sim (A_0+\partial_t\pi)^2 \,, \quad \tilde{F}_{\mu}\tilde{F}^{\mu} \sim (\partial_t A_i)^2 + (\partial_i A_0)^2
    \,,
    \label{kinetic_decoupe}
\end{align}
where the coefficients are omitted for simplicity of the notation. The gradient term of $A_i$ is generated by
\begin{align}
    \tilde{F}_{\mu\nu}\tilde{F}^{\mu\nu} \sim (\partial_i A_j)^2
    \,,
    \label{gradient_decoupe}
\end{align}
while the operators of the order of $(\partial_i \pi)^2$ can be obtained from
\begin{align}
    \delta \tilde{K} \,, ~ \delta \spatialR &\sim \partial_t [(\partial_i \pi)^2+A_i^2]\,, 
    \label{K_decouple}
    \\
    \delta \tilde{g}^{00}\delta \tilde{K}\,, ~\delta \tilde{g}^{00}\delta \spatialR &\sim A_0 \partial_i^2 \pi + \partial_t (\pi \partial_i^2 \pi) 
    \,. \label{gK_decouple}
\end{align}
In addition, we have
\begin{align}
    \tilde{\sigma}_{\mu\nu}\tilde{\sigma}^{\mu\nu}- \frac{2}{3} \delta \tilde{K}^2 &\sim (\partial_i A_j)^2\,, \\
    \delta \tilde{K}^2\,,~ \delta\tilde{K}\delta\spatialR\,, ~ \delta\spatialR^2 &\sim (\partial_i^2 \pi)^2
    \,,
\end{align}
where the former contributes to the gradient term of the transverse modes (and the kinetic term of the tensor modes) while the latter provide higher-order gradient terms of the longitudinal mode. Note that the operators~$\delta \tilde{K}$ and $\delta \spatialR$ may be comparable around the cosmological background despite the fact that $\tilde{K}$ is a first derivative term while $\spatialR$ is a second derivative. For instance, let us consider the following terms
\begin{align}
    \tilde{K}^2\,, \quad \spatialR\,,
    \label{K2+R}
\end{align}
both of which contain two derivatives. Perturbing the cosmological background, we obtain
\begin{align}
    \tilde{K}_{\rm BG}(t) \delta \tilde{K}\,, \quad \delta \spatialR
    \,,
\end{align}
where the first one is linear in the apparently ``first-order'' derivative term~$\delta \tilde{K}$ while the second one is the apparently ``second-order'' derivative term~$\delta \spatialR$. However, as clearly seen from \eqref{eqn:delta1Ktilde}--\eqref{eqn:delta1R3} and \eqref{eqn:delta2Ktilde}--\eqref{eqn:delta2R3}, these operators are of the same order. Since the EFT is a theory for the perturbations around a given background, one should count the number of the derivatives acting on the perturbations. Then, we should equally treat $\delta \tilde{K}$ and $\delta \spatialR$ around a general cosmological background. The operators in \eqref{kinetic_decoupe}, \eqref{gradient_decoupe}, \eqref{K_decouple}, and \eqref{gK_decouple} are regarded as the leading operators for the longitudinal and transverse modes of the EFT.

It is important to notice that the bare operator~$(\partial_i \pi)^2$ is prohibited in the present symmetry breaking pattern although $\partial_t [(\partial_i \pi)^2]$ and $A_0 \partial_i^2 \pi$ are allowed. In the scalar-tensor EFT, the operator~$\delta g^{00}$ is allowed and provides $(\partial_i \pi)^2$. On the other hand, in the present case the background equation of motion imposes $\barL_{\tilde{g}^{00}}=0$ and thus the term linear in $\delta\tilde{g}^{00}$ is forbidden. The gradient term of the longitudinal mode~$(\partial_i \pi)^2$ appears through integration by parts if the coefficient of $\partial_t [(\partial_i \pi)^2]$ is time-dependent or integrating out the non-dynamical variable~$A_0$ when $\delta \tilde{g}^{00}\delta \tilde{K}$ or $ \delta \tilde{g}^{00}\delta \spatialR $ is present; that is why we have regarded \eqref{K_decouple} and \eqref{gK_decouple} as the leading operators. However, there is a special case where $(\partial_i \pi)^2$ is totally prohibited, called the gauged ghost condensate~\cite{Cheng:2006us,Mukohyama:2006mm}. Let us consider the Minkowski background, $H=0$, in which $\delta \spatialR$ is irrelevant for the longitudinal mode. The operator~$\delta \tilde{K}\sim \partial_t [(\partial_i \pi)^2+A_i^2]$ is total derivative because the translation invariance of the Minkowski background requires a time-independent Taylor coefficient. The operator~$\delta \tilde{g}^{00}\delta \tilde{K}$ is prohibited when we additionally impose the time-reversal symmetry. As a result, we do not have \eqref{K_decouple} or \eqref{gK_decouple}, concluding the absence of $(\partial_i \pi)^2$. In this case, the leading operator for the gradient term of the longitudinal mode is given by $\delta \tilde{K}^2 \sim (\partial_i^2 \pi)^2$.\footnote{Nonetheless, the dispersion relation of the gauge ghost condensate takes the form~$\omega^2=c_s^2 k^2 + \alpha k^4/\cutoff^2$, rather than $\omega^2 = \alpha k^4/\cutoff^2$~\cite{Cheng:2006us,Mukohyama:2006mm}. The appearance of the $k^2$ term becomes manifest only after integrating out the non-dynamical variable $A_0$. For details, see Sec.~\ref{sec:STpert}.} Note that $\delta \spatialR$ is comparable to $\delta \tilde{K}$ around the background with $H\neq 0$, meaning that the operators~$\delta \tilde{K}\delta \spatialR$ and $\delta \spatialR^2$ are also the leading operators in the gauged ghost condensate with $H\neq 0$.

Other operators are sub-leading and/or non-linear in perturbations: the operators~$\tilde{\sigma}^{\mu\nu}\spatialR^T_{\mu\nu}$ and $\spatialR^T_{\mu\nu}\spatialR^T{}^{\mu\nu}$ lead to higher-derivative terms for the tensor modes and inclusion of the derivatives~$\pounds_{\tilde{n}}$ or $\tilde{D}_{\mu}$ increases the number of the derivatives. It is important to stress that we have assumed the decoupling limit in the analysis. In general, the metric perturbations cannot be ignored in the context of dark energy and, in such a situation, there can be additional operators as discussed in Appendix~\ref{sec:Lie}. 
Furthermore, in generic degenerate theories, called extended vector-tensor theories~\cite{Kimura:2016rzw}, the eigenstate of the longitudinal mode must be a mixture of the perturbations of the preferred vector and the metric, implying that the metric perturbations play an important role to identify the leading operators. In that case, we should first define the eigenstate and then count derivatives acting on it. For simplicity, however, we do not consider such generic degenerate theories in this paper.\footnote{By \emph{generic} degenerate theories, we mean those in which the Euler-Lagrange equations contain third- or higher-order derivatives but can be recomposed to yield second-order equations, and this recomposition amounts to the definition of the eigenstate for the longitudinal mode. Therefore, the generalized Proca theory~\cite{Heisenberg:2014rta}, for which the Euler-Lagrange equations are intrinsically second order, is within the scope of this paper.}
Then, the operators presented in \eqref{expansionL0} and \eqref{expansionL2} cover all the leading operators of the EFT around the general cosmological background and the gauged ghost condensate phase,  at least under the decoupling limit of gravity.

To be more explicit in counting the order of the operators, we assume the following typical orders for the Taylor coefficients
\begin{align}
    \begin{split}
    &\barL_0  = \mathcal{O}(\Mpl^2 H^2) \,, \quad \barL_{\tilde{K}} = \mathcal{O}(\Mpl^2  H) \,, \quad \barL_{\spatialR}=\mathcal{O}(\Mpl^2 )
    \,, \quad
    \barL_{\tilde{g}^{00}\tilde{g}^{00}} = \mathcal{O}(\Mpl^2 H^2) 
    \,, \\
    &\barL_{\tilde{g}^{00}\tilde{K}} = \mathcal{O}(\Mpl^2 H)
    \,, \quad
    \barL_{\tilde{K}\tilde{K}}= \mathcal{O}(\Mpl^2)
    \,, \quad
    \barL_{\tilde{\sigma}_{**}^2} = \mathcal{O}(\Mpl^2)
    \,, \quad
    \barL_{\tilde{g}^{00}\spatialR} = 
    \mathcal{O}(\Mpl^2)
    \,, \quad
    \\
    &\barL_{\tilde{F}_{*}^2} = \mathcal{O}(1)\,, \quad 
    \barL_{\tilde{F}_{**}^2} = \mathcal{O}(1)
    \,,
    \end{split}
    \label{scaling}
\end{align}
and
\begin{align}
    \barL_{\tilde{K}\tilde{K}} + \frac{4}{3}\barL_{\tilde{\sigma}_{**}^2} = \mathcal{O}(\Mpl^2 H^2 \cutoff^{-2})
    \,, \quad
    \barL_{\tilde{K}\spatialR} = \mathcal{O}(\Mpl^2 H \cutoff^{-2})
    \,, \quad
    \barL_{\spatialR \spatialR} = \mathcal{O}(\Mpl^2 \cutoff^{-2})\,,
    \label{subleading_scaling}
\end{align}
where $\Mpl$ is the Planck mass and $\cutoff$ is the cutoff scale of the EFT. One can easily confirm that, under the scaling~\eqref{scaling}, the leading operators equally contribute around the cosmological background. We assume that the higher-derivative operators are suppressed by the use of $\cutoff$; for instance, the coefficients of $\tilde{\sigma}^{\mu\nu}\spatialR^T_{\mu\nu}$ and $\spatialR^T_{\mu\nu} \spatialR^T{}^{\mu\nu}$ are supposed to be of the order of $\Mpl^2 \cutoff^{-1}$ and $\Mpl^2 \cutoff^{-2}$, respectively. Then, the higher-derivative operators can be ignored in the scales much lower than $\cutoff$. In the gauged ghost condensate, the operator~$(\partial_i \pi)^2$ is absent; then, the operators corresponding to \eqref{subleading_scaling} give the leading gradient terms of the order of $\frac{1}{\cutoff^2}(\partial_i^2 \pi)^2$.

The functions~$\mathcal{L}_0$, $\mathcal{L}_{\tilde{F}_{*}^2}$, $\mathcal{L}_{\tilde{F}_{**}^2}$, and $\mathcal{L}_{\tilde{\sigma}_{**}^2}$ are invariant under the residual symmetries of the EFT while the Taylor coefficients, the functions with the bar, are not invariant due to the absence of $t$ in the building blocks. The functions with the bar are functions of the background part of \eqref{pert_non_scalars} and thus implicit functions of the time. Using the chain rule, we obtain
\begin{align}
    \frac{\D}{\D t}\barL_0(t)&=\left. \frac{\D}{\D t} \mathcal{L}_0(\tilde{g}^{00},\tilde{K},\spatialR,\pounds_{\tilde{n}})\right|_{\rm BG} 
    \nn
    &= \barL_{\tilde{g}^{00}}(t) \frac{\D}{\D t}\tilde{g}^{00}_{\rm BG}(t) + \barL_{\tilde{K}}(t) \frac{\D}{\D t}\tilde{K}_{\rm BG}(t) + \barL_{\spatialR}(t) \frac{\D}{\D t}\spatialR_{\rm BG}(t) + \cdots
    \,,
    \label{consistency1}
\end{align}
and
\begin{align}
    \frac{\D}{\D t}\barL_{\tilde{g}^{00}}(t) & = \barL_{\tilde{g}^{00}\tilde{g}^{00}}(t) \frac{\D}{\D t}\tilde{g}^{00}_{\rm BG}(t) + \barL_{\tilde{g}^{00}\tilde{K}}(t) \frac{\D}{\D t}\tilde{K}_{\rm BG}(t) + \barL_{\tilde{g}^{00}\spatialR}(t) \frac{\D}{\D t}\spatialR_{\rm BG}(t) + \cdots
    \,, \label{consistency2} \\
    \frac{\D}{\D t}\barL_{\tilde{K}}(t) & = \barL_{\tilde{g}^{00}\tilde{K}}(t) \frac{\D}{\D t}\tilde{g}^{00}_{\rm BG}(t) + \barL_{\tilde{K}\tilde{K}}(t) \frac{\D}{\D t}\tilde{K}_{\rm BG}(t)  + \cdots
    \,, \\
    \frac{\D}{\D t} \barL_{\spatialR}(t) &= \barL_{\tilde{g}^{00}\spatialR}(t) \frac{\D}{\D t}\tilde{g}^{00}_{\rm BG}(t) + \cdots
    \,,
    \label{consistency4}
\end{align}
where the ellipses stand for the terms suppressed by $\cutoff$. Here, \eqref{scaling} and  \eqref{subleading_scaling} are used and the Lie derivatives are supposed to be suppressed by $\cutoff$. Equations~\eqref{consistency1}--\eqref{consistency4} provide the consistency relations between the Taylor coefficients through the background dynamics. One can also derive more consistency relations between higher-order coefficients.

\subsection{EFT action}\label{subsec-EFT-action}
We have introduced the essential ingredients for the formulation of the vector-tensor EFT in subsections~\ref{sec:symmetry}--\ref{sec:expansion}. In this subsection, we reorganize the Lagrangian to be more compatible with the conventional EFT of inflation/dark energy.

Let us formulate the EFT action in terms of the perturbations around the homogeneous and isotropic background configuration
\begin{align}
        \bar{g}_{\mu\nu}\D x^{\mu}\D x^{\nu}=-\bar{N}^2(t)\D t^2+a^2(t)\gamma_{ij}\D x^i \D x^j
        \,, \quad \bar{A}_{\mu}=(\bar{A}_0(t), 0)\,,
\end{align}
where $\gamma_{ij}$ is the spatial metric for the maximally symmetric space. Although there is a freedom of the time reparametrization, $t\to t'(t)$, which can be used to set either $\bar{N}=1$ or $\bar{A}_0=0$, we keep the general expressions for the background quantities. Based on the discussions in the previous subsections, the leading-order action of the EFT can be rewritten as
\begin{align}
    S&=\int \D^4x \sqrt{-g}\mathcal{L}_{\rm DE} + S_{\rm m}
    \,, \nn
    \mathcal{L}_{\rm DE}&= \frac{\Mpl^2}{2}f(t) \tilde{R} - \Lambda(t)-c(t) \tilde{g}^{00}-d(t)\tilde{K} + \mathcal{L}_{\rm DE}^{(2)}
    \,,\nn
    \mathcal{L}_{\rm DE}^{(2)}&=
    \frac{1}{2} M_2^4(t) \left(\frac{\delta\tilde{g}^{00}}{-\tilde{g}_{\rm BG}^{00} }\right)^2
    - \frac{1}{2} {\bar M}_1^3(t) \left(\frac{\delta\tilde{g}^{00}}{-\tilde{g}_{\rm BG}^{00} }\right) \delta {\tilde K}
    - \frac{1}{2} {\bar M}_2^2(t) \delta {\tilde K}^2
    - \frac{1}{2} {\bar M}_3^2(t) \delta \tilde{K}^{\alpha}{}_{\beta} \delta \tilde{K}^{\beta}{}_{\alpha} 
    \nonumber \\ 
    & \quad + \frac{1}{2}\mu_1^2(t)\left(\frac{\delta\tilde{g}^{00}}{-\tilde{g}_{\rm BG}^{00} }\right)\delta \spatialR 
    - \frac{1}{4} \gamma_1(t) F_{\alpha\beta} F^{\alpha\beta}  
    - \frac{1}{4} \gamma_2(t) {\tilde F}_{\alpha\beta} {\tilde F}^{\alpha\beta} 
    \nn
    & \quad +\frac{1}{2}\bar{M}_{4}(t) \delta \tilde{K}  \delta \spatialR
    +\frac{1}{2}\lambda_{1}(t) \delta \spatialR^2
    + \cdots \,,
        \label{EFT_L}
\end{align}
where
\begin{align}
    \tilde{R} & \equiv  {}^{(3)}\! \tilde{R} + \tilde{K}_{\mu\nu}\tilde{K}^{\mu\nu}-\tilde{\omega}_{\mu\nu}\tilde{\omega}^{\mu\nu}-\tilde{K}^2 
    =R + 2\nabla_{\mu}(\tilde{a}^{\mu}-\tilde{K} \tilde{n}^{\mu})
    \,, \\
 \delta \tilde{K}_{\mu\nu} & \equiv \tilde{\sigma}_{\mu\nu}+\frac{1}{3}\delta \tilde{K} \tilde{h}_{\mu\nu} =\tilde{K}_{\mu\nu}-\frac{1}{3}\tilde{K}_{\rm BG}(t)\tilde{h}_{\mu\nu}
    \,,
\label{def_deltaK}
\end{align}
which are the four-dimensional Ricci scalar with the divergence term subtracted and the perturbation of the expansion tensor, respectively. 
The Lagrangian~\eqref{EFT_L} takes a similar form as the conventional EFT of inflation/dark energy~\cite{Cheung:2007st,Gubitosi:2012hu} and a comparison will be discussed shortly.

The coefficients in \eqref{EFT_L} and the Taylor coefficients appearing in \eqref{expansionL0} and \eqref{expansionL2} are related via
\begin{align}
    \begin{split}
    & \Mpl^2 f = 2\barL_{\spatialR} \,, \quad 
    \Lambda = - \barL_0 + \barL_{\tilde{g}^{00}} \tilde{g}^{00} _{\rm BG} + \barL_{\tilde{K}} \tilde{K}_{\rm BG}
    +\barL_{\spatialR}\left( \frac{2}{3}\tilde{K}_{\rm BG}^2 + \spatialR_{\rm BG} \right)
    \,, \\
    &c= -\barL_{\tilde{g}^{00}}\,, \quad
    d=-\barL_{\tilde{K}}-\frac{4}{3}\barL_{\spatialR}\tilde{K}_{\rm BG}
    \,, \quad
    M_2^4=\barL_{\tilde{g}^{00}\tilde{g}^{00}}(\tilde{g}^{00}_{\rm BG})^2
    \,, \\
    &\bar{M}_1^3=2\barL_{\tilde{g}^{00}\tilde{K}}\tilde{g}^{00}_{\rm BG}
    \,, \quad
    \bar{M}_2^2=-2\barL_{\spatialR}-\barL_{\tilde{K}\tilde{K}}+\frac{2}{3}\barL_{\tilde{\sigma}_{**}^2}
    \,, \quad
    \bar{M}_3^2=2(\barL_{\spatialR}-\barL_{\tilde{\sigma}_{**}^2})
    \,, \\
    &\mu_1^2=-2\barL_{\tilde{g}^{00}\spatialR} \tilde{g}^{00}_{\rm BG}\,, \quad
    \gamma_1=2\barL_{\tilde{F}_{*}^2}
    \,, \quad
    \gamma_2=-4\barL_{\tilde{F}_{**}^2}-2\barL_{\tilde{F}_{*}^2} + \barL_{\spatialR} \frac{\g^2}{\tilde{g}^{00}_{\rm BG} }
    \,, \\
    &\bar{M}_4=2\barL_{\tilde{K}\spatialR}
    \,, \quad
    \lambda_1 = \barL_{\spatialR\spatialR}
    \,.
    \end{split}
\end{align}
Equations~\eqref{scaling} and \eqref{subleading_scaling} then read
\begin{align}
    \begin{split}
    &f=\mathcal{O}(1)\,, \quad \Lambda = \mathcal{O}(\Mpl^2 H^2)\,, \quad d=\mathcal{O}(\Mpl^2 H)\,, \quad M_2^4=\mathcal{O}(\Mpl^2 H^2)
    \,, \quad \bar{M}_1^3=\mathcal{O}(\Mpl^2 H)
    \,, \\
    &\bar{M}_2^2=\mathcal{O}(\Mpl^2)\,, \quad \bar{M}_3^2=\mathcal{O}(\Mpl^2)
    \,, \quad 
    \mu_1^2 = \mathcal{O}(\Mpl^2)\,, \quad
    \gamma_1=\mathcal{O}(1)\,, \quad \gamma_2=\mathcal{O}(1)
    \,,
    \end{split}
    \label{scalingM}
\end{align}
and
\begin{align}
    \bar{M}_2^2+\bar{M}_3^2=\mathcal{O}(\Mpl^2 H^2 \cutoff^{-2})
    \,, \quad 
    \bar{M}_4 = \mathcal{O}(\Mpl^2 H \cutoff^{-2})
    \,, \quad
    \lambda_1 = \mathcal{O}(\Mpl^2 H^2 \cutoff^{-2})
    \,.
    \label{subleading_scalingM}
\end{align}
We also find $c={\cal O}(\Mpl^2 H^2)$ but we did not write it in \eqref{scalingM} because, as we will see in subsection~\ref{sec:background}, one of the background equations of motion imposes $c=0$ without using the decoupling limit. Note that the higher-derivative operators, $\delta \tilde{K}  \delta \spatialR$ and $\delta \spatialR^2 $, are added to the leading-order Lagrangian since they can be leading ones around the background of the gauged ghost condensate. If one is interested in a general background away from the gauged ghost condensate, one may ignore $\delta \tilde{K}  \delta \spatialR$ and $\delta \spatialR^2 $ and can assume $\bar{M}_2^2+\bar{M}_3^2\simeq 0 $ at the leading order. In terms of the new EFT coefficients defined in \eqref{EFT_L}, the consistency relations~\eqref{consistency1}--\eqref{consistency4} are rewritten as
\begin{align}
    \dot{\Lambda}+3H(\dot{d}+\Mpl^2\dot{f} H) - \frac{1}{2}\Mpl^2 \dot{f} \spatialR_{\rm BG} + \dot{c} \tilde{g}^{00}_{\rm BG} &\simeq 0
    \,, \label{consistency1M} \\
    2M_2^4 \frac{\D}{\bar{N} \D t} \ln (-\tilde{g}^{00}_{\rm BG}) + 3\bar{M}_1^3 \dot{H} - \mu_1^2 \frac{\D}{\bar{N} \D t} \spatialR_{\rm BG} + 2 \dot{c} \tilde{g}^{00}_{\rm BG}
    &\simeq 0
    \,, \label{consistency2M}\\
    \dot{d}+2\Mpl^2 \dot{f} H-3\dot{H}\left(\bar{M}_2^2 + \frac{1}{3}\bar{M}_3^2\right) + \frac{1}{2}\bar{M}_1^3 \frac{\D}{\bar{N} \D t} \ln (-\tilde{g}^{00}_{\rm BG})
    &\simeq 0
    \,, \label{consistency3M} \\
    \Mpl^2 \dot{f} + \mu_1^2 \frac{\D}{\bar{N} \D t} \ln (-\tilde{g}^{00}_{\rm BG}) &\simeq 0
    \,, \label{consistency4M}
\end{align}
where a dot is the derivative with respect to the cosmic time and we have used $\tilde{K}_{\rm BG}=3H$. The consistency relations~\eqref{consistency1M}--\eqref{consistency4M} are off-shell relations; that is, they have to hold without using the equations of motion. Recall that \eqref{consistency1M}--\eqref{consistency4M} are derived by ignoring the operators suppressed by the cutoff scale under \eqref{scalingM} and \eqref{subleading_scalingM}, so $\bar{M}_2^2 + \frac{1}{3}\bar{M}_3^2$ in \eqref{consistency3M} should be understood as $\bar{M}_2^2 + \frac{1}{3}\bar{M}_3^2\simeq \frac{2}{3}\bar{M}_2^2$.

It is straightforward to show that the EFT action~\eqref{EFT_L} is invariant under the residual symmetries when the consistency relations~\eqref{consistency1M}--\eqref{consistency4M} are imposed and vice versa. Since the invariance under the spatial diffs is manifest, we only consider the invariance under the combined $U(1)$ and time diffs,
\begin{align}
    t \to t - \g \chi\,, \quad A_{\mu}\to A_{\mu}+\partial_{\mu}\chi
    \,. \label{U(1)_time}
\end{align}
The building blocks~\eqref{background_blocks} and \eqref{pert_blocks} are constructed so that they respect the symmetry~\eqref{U(1)_time} while the perturbations~\eqref{pert_non_scalars} and the time-dependent coefficients in \eqref{EFT_L} are not invariant quantities; for example, $\delta \tilde{g}^{00}=\tilde{g}^{00}-\tilde{g}^{00}_{\rm BG}(t)$ is transformed as
\begin{align}
    \delta \tilde{g}^{00} \to \delta \tilde{g}^{00} + \frac{\D\tilde{g}^{00}_{\rm BG}(t)}{\D t} \g \chi\,,
\end{align}
under the infinitesimal gauge transformation. The variation of the action under the infinitesimal transformation is given by the form
\begin{align}
    \delta S = \int \D^4x \sqrt{-g}\, \g \chi \left[\delta \mathcal{L}(t) + \delta \mathcal{L}_{\delta \tilde{g}^{00}}(t) \delta \tilde{g}^{00} + \delta \mathcal{L}_{\delta \tilde{K}}(t) \delta \tilde{K} + \delta \mathcal{L}_{\delta \spatialR}(t) \delta \spatialR + \cdots \right]
    \,,
    \label{deltaS}
\end{align}
where the ellipsis stands for terms which are quadratic or higher order in perturbations.
Since $\delta S = 0$ has to hold for any configurations of the fields, the invariance under \eqref{U(1)_time} requires
\begin{align}
    \delta \mathcal{L}&= 
     \delta \mathcal{L}_{\delta \tilde{g}^{00}} =
     \delta \mathcal{L}_{\delta \tilde{K}} =
     \delta \mathcal{L}_{\delta \spatialR} =0\,,
\end{align}
which agrees with \eqref{consistency1M}--\eqref{consistency4M} up to the terms suppressed by $\cutoff$.

In this way, the residual symmetry~\eqref{U(1)_time} is implemented as the consistency relations between the coefficients. One may first regard the coefficients in \eqref{EFT_L} as independent time-dependent functions during intermediate steps of calculations and impose the consistency relations \eqref{consistency1M}--\eqref{consistency4M} at the final step. The advantage of this method becomes clear when one compares the vector-tensor theories with the scalar-tensor theories. Let us set $\g=0$ without imposing the consistency relations~\eqref{consistency1M}--\eqref{consistency4M}. The EFT action~\eqref{EFT_L} reduces to the scalar-tensor EFT with a gauge field~$A_{\mu}$ in which the equation of motion admits a trivial solution~$A_{\mu}=0$. Then, all predictions of the scalar-tensor EFT can be obtained if all EFT coefficients are considered as independent functions. We next consider the continuous limit~$\g \to 0$ under the consistency relations~\eqref{consistency1M}--\eqref{consistency4M}. One can easily show that imposing \eqref{consistency1M}--\eqref{consistency4M} with $\g\to 0$ is equivalent to demanding that the action is invariant under a global change of the time coordinate, $t \to t + \chi_0$ with a constant~$\chi_0$. Since the St\"{u}ckelberg field $\tilde{t}$ has been identified with the time coordinate, this global symmetry is recast as the global shift symmetry of the scalar field when the St\"{u}ckelberg field is reintroduced. Physically, the parameter~$\g$ can be interpreted as the gauge coupling associated with the shift symmetry, so the localized (gauged) shift symmetry reduces to the global shift symmetry in the limit~$\g \to 0$. Therefore, the action~\eqref{EFT_L} includes three different classes of the models depending on whether $\g=0$ or $\g\neq 0$ and whether or not \eqref{consistency1M}--\eqref{consistency4M} are imposed. Using the EFT action~\eqref{EFT_L}, we will elaborate a unified formalism of the linear perturbations in Sec.~\ref{sec:comparison}. However, before doing that, let us clarify how a particular vector-tensor theory can be realized in our EFT framework in the next section.

\section{Dictionary}
\label{sec:dictionary}

As a simple demonstration, let us consider a subclass of the generalized Proca theory~\cite{Heisenberg:2014rta} described by the following action
    \be
    S_{\rm GP}=\int \D^4x\sqrt{-g}\brc{G_2(X,F,Y)+G_3(X)\nabla_\mu A^\mu+G_4(X)R+G_{4X}\brb{\brp{\nabla_\mu A^\mu}^2-\nabla_\mu A_\nu \nabla^\nu A^\mu}}\,, \label{GP}
    \ee
where $G_2$ is an arbitrary function of $(X,F,Y)$, $G_3$ and $G_4$ are arbitrary functions of $X$, and a subscript~$X$ denotes the derivative with respect to $X$, with
    \be
    X\equiv -\frac{1}{2}A_\mu A^\mu\,, \quad
    F\equiv -\frac{1}{4}F_{\mu\nu}F^{\mu\nu}\,, \quad
    Y\equiv A^\mu A^\nu F_{\mu\alpha}F_{\nu}{}^{\alpha}\,.
    \ee
The $U(1)$ gauge symmetry can be restored by introducing a St{\"u}ckelberg field~$\tilde{t}$ as $A_\mu\to A_\mu+\g^{-1}\partial_\mu\tilde{t}$.
We then take the unitary gauge where the time coordinate~$t$ coincides with $\tilde{t}$.
Now, the residual symmetries are the spatial diffs and a combination of the $U(1)$ and time diffs, on which our vector-tensor EFT was constructed.
Since $A_\mu$ is mapped to
    \be
    A_\mu\to \g^{-1}\tilde{\delta}^0{}_\mu
    =-\frac{\sqrt{-\tilde{g}^{00}}}{\g}\tilde{n}_\mu \,,
    \ee
in the unitary gauge, one can evaluate the quantities~$X$ and $Y$ as
    \be
    X=-\frac{\tilde{g}^{00}}{2\g^2}\,, \quad
    Y=-\frac{\tilde{g}^{00}}{\g^2}\tilde{F}_\mu\tilde{F}^\mu
    =\frac{\tilde{g}^{00}}{2\g^2}\brp{F_{\mu\nu}F^{\mu\nu}-\tilde{F}_{\mu\nu}\tilde{F}^{\mu\nu}}\,.
    \ee
Note that $F$ and $Y$ are at least quadratic order in perturbations on a homogeneous and isotropic background. Using $\nabla_\mu A_\nu = - \nabla_\mu (\sqrt{2X} {\tilde n}_\nu)$ and performing integration by parts, the action~\eqref{GP} can be written in terms of the geometrical quantities introduced in Sec.~\ref{sec:building_blocks} as follows:
    \begin{align}
    S_{\rm GP}&=\int \D^4x\sqrt{-g}\brb{G_2+F_{3} \tilde{K}+G_4\spatialR
    +\brp{G_4-2XG_{4X}}\brp{\tilde{K}_{\mu\nu}\tilde{K}^{\mu\nu}-\tilde{K}^2-\tilde{\omega}_{\mu\nu}\tilde{\omega}^{\mu\nu}}} \nonumber \\
    &=\int \D^4x\sqrt{-g}\brb{G_2+F_{3} \tilde{K}+G_4\tilde{R}
    -2XG_{4X}\brp{\tilde{K}_{\mu\nu}\tilde{K}^{\mu\nu}-\tilde{K}^2-\tilde{\omega}_{\mu\nu}\tilde{\omega}^{\mu\nu}}}\,,
    \end{align}
where we have defined the function~$F_3(X)$ so that $F_{3X}=-(2X)^{1/2}G_{3X}$.
Expanding each term in the Lagrangian around the cosmological background, we have
    \be
    \begin{split}
    G_2&=\bar{G}_{2}-\bar{X}\bar{G}_{2X}-\frac{\bar{G}_{2X}}{2\g^2}\tilde{g}^{00}+\frac{\bar{G}_{2XX}}{8\g^4}\brp{\delta\tilde{g}^{00}}^2 
    -\brp{\frac{\bar{G}_{2F}}{4}+\bar{X}\bar{G}_{2Y}}F_{\mu\nu}F^{\mu\nu} \\
    &\quad+\bar{X}\bar{G}_{2Y}\tilde{F}_{\mu\nu}\tilde{F}^{\mu\nu}+\cdots\,, \\
    F_3\tilde{K}&=-\bar{X}\bar{F}_{3X}\tilde{K}_{\rm BG}-\frac{\bar{F}_{3X}\tilde{K}_{\rm BG}}{2\g^2}\tilde{g}^{00}+\bar{F}_3\tilde{K}+\frac{\bar{F}_{3XX}\tilde{K}_{\rm BG}}{8\g^4}\brp{\delta\tilde{g}^{00}}^2-\frac{\bar{F}_{3X}}{2\g^2}\delta\tilde{g}^{00}\delta\tilde{K}+\cdots\,, \\
    G_4\tilde{R}&=\bar{G}_4\tilde{R}-\bar{X}\bar{G}_{4X}\tilde{R}_{\rm BG}-\frac{\bar{G}_{4X}\tilde{R}_{\rm BG}}{2\g^2}\tilde{g}^{00}+\frac{\bar{G}_{4XX}\tilde{R}_{\rm BG}}{8\g^4}\brp{\delta\tilde{g}^{00}}^2+\frac{2\bar{G}_{4X}\tilde{K}_{\rm BG}}{3\g^2}\delta\tilde{g}^{00}\delta\tilde{K} \\
    &\quad-\frac{\bar{G}_{4X}}{2\g^2}\delta\tilde{g}^{00}\delta\spatialR+\cdots\,,
    \end{split}
    \ee
with $\tilde{R}_{\rm BG}\equiv \spatialR_{\rm BG}-(2/3)\tilde{K}_{\rm BG}^2$ being the background value of $\tilde{R}$, and
    \begin{align}
    &-2XG_{4X}\brp{\tilde{K}_{\mu\nu}\tilde{K}^{\mu\nu}-\tilde{K}^2-\tilde{\omega}_{\mu\nu}\tilde{\omega}^{\mu\nu}} \nonumber \\
    &=-\frac{4}{3}\tilde{K}_{\rm BG}^2\bar{X}(2\bar{G}_{4X}+\bar{X}\bar{G}_{4XX})-\frac{2\tilde{K}_{\rm BG}^2}{3\g^2}(\bar{G}_{4X}+\bar{X}\bar{G}_{4XX})\tilde{g}^{00}+\frac{8}{3}\bar{X}\bar{G}_{4X}\tilde{K}_{\rm BG}\tilde{K} \nonumber \\
    &\quad +\frac{\tilde{K}_{\rm BG}^2}{6\g^4}(2\bar{G}_{4XX}+\bar{X}\bar{G}_{4XXX})\brp{\delta\tilde{g}^{00}}^2-\frac{4\tilde{K}_{\rm BG}}{3\g^2}(\bar{G}_{4X}+\bar{X}\bar{G}_{4XX})\delta\tilde{g}^{00}\delta\tilde{K} \nonumber \\
    &\quad -2\bar{X}\bar{G}_{4X}\brp{\delta\tilde{K}_{\mu\nu}\delta\tilde{K}^{\mu\nu}-\delta\tilde{K}^2}+\frac{\bar{G}_{4X}}{4}\tilde{F}_{\mu\nu}\tilde{F}^{\mu\nu}+\cdots\,,
    \end{align}
where the barred functions are evaluated at the background, i.e.~$(X,F,Y)=(\bar{X},0,0)$ with $\bar{X}\equiv -\tilde{g}^{00}_{\rm BG}(t)/2\g^2$.
Hence, the generalized Proca theory can be written in the form of \eqref{EFT_L} with the following coefficients:
    \be
    \begin{split}
    &\Mpl^2 f=2\bar{G}_4\,, \quad
    d=-\bar{F}_3-\frac{8}{3}\bar{X}\bar{G}_{4X}\tilde{K}_{\rm BG}\,, \\
    &\Lambda=-\bar{G}_{2}+\bar{X}\bar{G}_{2X}+\bar{X}\bar{F}_{3X}\tilde{K}_{\rm BG}+\bar{X}\bar{G}_{4X}\tilde{R}_{\rm BG}+\frac{4}{3}\tilde{K}_{\rm BG}^2\bar{X}(2\bar{G}_{4X}+\bar{X}\bar{G}_{4XX})\,, \\
    &c=
    \frac{\bar{X}}{-\tilde{g}^{00}_{\rm BG}}\brb{\bar{G}_{2X}+\bar{F}_{3X}\tilde{K}_{\rm BG}+\bar{G}_{4X}\tilde{R}_{\rm BG}+\frac{4}{3}\tilde{K}_{\rm BG}^2(\bar{G}_{4X}+\bar{X}\bar{G}_{4XX})}\,, \\
    &M_2^4=\bar{X}^2\brb{\bar{G}_{2XX}+\bar{F}_{3XX}\tilde{K}_{\rm BG}+\bar{G}_{4XX}\tilde{R}_{\rm BG}+\frac{4}{3}\tilde{K}_{\rm BG}^2(2\bar{G}_{4XX}+\bar{X}\bar{G}_{4XXX})}\,, \\
    &\bar{M}_1^3=2\bar{X}\brb{\bar{F}_{3X}+\frac{4}{3}\tilde{K}_{\rm BG}(\bar{G}_{4X}+2\bar{X}\bar{G}_{4XX})}\,, \quad
    \bar{M}_2^2=-\bar{M}_3^2=-4\bar{X}\bar{G}_{4X}\,, \\
    &\mu_1^2=-2\bar{X}\bar{G}_{4X}\,, \quad
    \gamma_1=\bar{G}_{2F}+4\bar{X}\bar{G}_{2Y}\,, \quad
    \gamma_2=-4\bar{X}\bar{G}_{2Y}-\bar{G}_{4X}\,,
    \quad
    \bar{M}_4=\lambda_1=0\,.
    \end{split}
    \label{dictionary_GP}
    \ee
It is easy to show that the EFT coefficients~\eqref{dictionary_GP} indeed satisfy the consistency relations~\eqref{consistency1M}--\eqref{consistency4M}.

\section{Comparison with scalar-tensor EFT}
\label{sec:comparison}

We study the spacetime without vector-type perturbations to see the difference from the conventional EFT of inflation/dark energy, namely the scalar-tensor EFT. As we explained in Sec.~\ref{sec:building_blocks}, when the vector perturbations are absent, we can choose the gauge so that the vector field~$A_{\mu}$ is given by
\begin{align}
    A_{\mu}=(A_0,0)\,, \quad A_0=\bar{A}_0(t)+\delta A_0(t,x^i)
    \,,
    \label{A_for_ST}
\end{align}
where the time coordinate is completely fixed except the freedom of the reparametrization, $t\to t'=t'(t)$. In this gauge choice, as explained at the end of subsection~\ref{sec:building_blocks}, $\tilde{K}_{\mu\nu}$ and $\spatialR_{\mu\nu\rho\sigma}$ coincide with the extrinsic curvature~$K_{\mu\nu}$ and the intrinsic curvature~$\spatialRST_{\mu\nu\rho\sigma}$ of the constant-time hypersurface, respectively. Moreover, ${\tilde F}_{\mu\nu}$ vanishes for \eqref{A_for_ST}. Then, the EFT action \eqref{EFT_L} is simplified to be
\begin{align}
    S&=\int \D^4x \sqrt{-g} \Bigg[
    \frac{\Mpl^2}{2}f(t) \left( \spatialRST+K_{\mu\nu}K^{\mu\nu}-K^2 \right) - \Lambda(t)-c(t) \tilde{g}^{00}-d(t)K 
 + \mathcal{L}_{\rm DE}^{(2)} 
    \Bigg] + S_{\rm m}
    \,, \label{EFT_for_ST}
 \nn
\mathcal{L}_{\rm DE}^{(2)} 
 &= 
    \frac{1}{2} M_2^4(t) \left(\frac{\delta\tilde{g}^{00}}{-\tilde{g}_{\rm BG}^{00} }\right)^2
- \frac{1}{2} {\bar M}_1^3(t) \left(\frac{\delta\tilde{g}^{00}}{-\tilde{g}_{\rm BG}^{00} }\right) \delta  K
- \frac{1}{2} {\bar M}_2^2(t) \delta K^2
- \frac{1}{2} {\bar M}_3^2(t) \delta K^{\alpha}{}_{\beta} \delta K^{\beta}{}_{\alpha} 
\nn
& 
+ \frac{1}{2}\mu_1^2(t)\left(\frac{\delta\tilde{g}^{00}}{-\tilde{g}_{\rm BG}^{00} }\right) \delta \spatialRST - \frac{1}{4} \gamma_1(t) F_{\alpha\beta} F^{\alpha\beta}  
+\frac{1}{2}\bar{M}_{4}(t) \delta K  \delta \spatialRST
    +\frac{1}{2}\lambda_{1}(t) \delta \spatialRST^2
+ \cdots\,,
\end{align}
where, in this particular ansatz, the norm of the preferred vector is
\begin{align}
    \tilde{g}^{00}=g^{00}(1+\g A_0)^2
    \,. 
    \label{tildeg00_ST}
\end{align}
The action~\eqref{EFT_for_ST} can be used to study the background dynamics, the scalar perturbations, and the tensor perturbations of the vector-tensor EFT and we shall focus on it in this section. We emphasize that the utility of \eqref{EFT_for_ST} is not limited to linear perturbations. The analysis can be systematically extended to non-linear orders by adding necessary operators, say $(\delta \tilde{g}^{00})^3$, and by considering consistency conditions unless the coupling to the vector perturbations~(vorticity) is concerned. In particular, for an inflationary setup when the vector perturbations usually decay, one can use the action~\eqref{EFT_for_ST} in a similar way to the one in the EFT of inflation~\cite{Cheung:2007st} and the main difference is that the consistency conditions~\eqref{consistency1M}--\eqref{consistency4M} should be imposed. On the other hand, one should go back to the action~\eqref{EFT_L} to study the vector perturbations, which we will separately discuss in Sec.~\ref{sec:vector_pert}.

\subsection{Background equations of motion}
\label{sec:background}

For simplicity, we consider the flat FLRW background
\begin{align}\label{FLRW-BG}
    \D\bar{s}^2=-\bar{N}^2(t) \D t^2 + a^2(t) \delta_{ij}\D x^i \D x^j
    \,.
\end{align}
The background equations of motion are obtained from the action~\eqref{EFT_for_ST} by demanding that the terms linear in the perturbations vanish. In \eqref{EFT_for_ST}, $\mathcal{L}_{\rm DE}^{(2)}$ is at least quadratic in the perturbations and thus does not contribute to the background equation of motion. 

As we have mentioned, the EFT coefficient~$c(t)$ has to vanish due to the background equation. As one can see from \eqref{EFT_for_ST}, the temporal component of the vector field~$A_0$ appears only through $\tilde{g}^{00}$ in the background part of the EFT action. Therefore, demanding the term linear in $\delta A_0$ to vanish, one obtains 
    \be
    \g(1+\g \bar{A}_0)c(t)=0 \,. \label{c_eq}
    \ee
This implies either $c=0$ or $1+\g \bar{A}_0=0$, the latter of which contradicts our input of the EFT, namely, the existence of the preferred vector at the level of the background. The equation~$c(t)=c(\tilde{g}^{00}_{\rm BG},\tilde{K}_{\rm BG})=0$ may be understood as the constraint that determines the background value of the norm of the preferred vector. 
Indeed, the tadpole cancellation condition~$c=0$ is consistent with the equation of motion for the vector field on a cosmological background in the generalized Proca theory obtained in~\cite{DeFelice:2016yws}. After performing integration by parts and using $c=0$, the background part of the EFT action~\eqref{EFT_for_ST} is simplified to
\begin{align}
    S=\int \D^4x \sqrt{-g} \Bigg[
    &\frac{\Mpl^2}{2}f(t) \left( \spatialRST+K_{\mu\nu}K^{\mu\nu}-K^2 \right) -\Lambda(t) + \dot{d}(t) \sqrt{ 1-\bar{N}^2\delta g^{00} } 
     + \mathcal{L}_{\rm DE}^{(2)} \Bigg] + S_{\rm m}
     \,,
    \label{LBG}
\end{align}
where we have defined
\begin{align}
    \delta g^{00} \equiv g^{00}+\bar{N}^{-2}
    \,.
\end{align}
The functions~$\Lambda(t)$ and $\dot{d}(t)$ are determined by
\begin{align}
\dot{d}(t)&= - \bar{\rho}_{\rm m} - \bar{p}_{\rm m} - 2\Mpl^2(  f \dot{H} + \dot{f} H )
\,, \label{d_eq} \\
\Lambda(t)&=-\bar{\rho}_{\rm m}+3 \Mpl^2  f H^2
\,, \label{lambda_eq}
\end{align}
and the matter field is subject to the conservation equation,
\begin{align}
\dot{\bar{\rho}}_{\rm m} +3H (\bar{\rho}_{\rm m} +\bar{p}_{\rm m})=0 \,.
\label{conserve}
\end{align}

In terms of the Taylor coefficients, the background equations of motion~\eqref{c_eq} and \eqref{lambda_eq} can be rewritten as
\begin{align}
    \barL_{\tilde{g}^{00}}(\tilde{g}^{00}_{\rm BG},\tilde{K}_{\rm BG})&=0
    \,, 
    \label{const_g00} \\
    \barL_0(\tilde{g}^{00}_{\rm BG},\tilde{K}_{\rm BG})-\barL_{\tilde{K}}(\tilde{g}^{00}_{\rm BG},\tilde{K}_{\rm BG})\tilde{K}_{\rm BG} &= \bar{\rho}_{\rm m}
    \label{Fri_L}
    \,.
\end{align}
These equations can be directly derived from the covariant action~\eqref{EFT_general_covariant} under the ansatz~\eqref{FLRW-BG} at the leading order of the derivative expansion. The solutions~$\tilde{g}^{00}_{\rm BG}=\tilde{g}^{00}_{\rm BG}(\bar{\rho}(t))$ and $\tilde{K}_{\rm BG}=\tilde{K}_{\rm BG}(\bar{\rho}(t))$ are uniquely found at least locally if and only if
\begin{align}
    {\rm det} 
    \begin{pmatrix}
    \barL_{\tilde{g}^{00}\tilde{g}^{00}} & \barL_{\tilde{g}^{00}\tilde{K}} \\
    \barL_{\tilde{g}^{00}\tilde{K}} & \barL_{\tilde{K}\tilde{K} }
    \end{pmatrix}
    =\barL_{\tilde{g}^{00}\tilde{g}^{00}}\barL_{\tilde{K}\tilde{K} }-(\barL_{\tilde{g}^{00}\tilde{K}})^2 \neq 0
    \,.
    \label{solvable_con}
\end{align}
This is the necessary and sufficient condition to uniquely determine the background evolution of the universe in the vector-tensor EFT at the leading order of the derivative expansion. In particular, the solvability condition~\eqref{solvable_con} concludes that the vacuum solution has to be $\tilde{g}^{00}_{\rm BG}={\rm constant}$ and $\tilde{K}_{\rm BG}=3H={\rm constant}$ in the vector-tensor EFT.

In the following analysis, however, we shall assume a stronger condition, $\barL_{\tilde{g}^{00}\tilde{g}^{00}} \neq 0$, in order that the background is uniquely determined even when gravity is decoupled. The condition~$\barL_{\tilde{g}^{00}\tilde{g}^{00}} \neq 0$ guarantees that the equation~$c=-\barL_{\tilde{g}^{00}}=0$ is solvable in terms of $\tilde{g}^{00}$ without any reference to the Friedmann equation~\eqref{Fri_L}. We then correctly interpret $c=-\barL_{\tilde{g}^{00}}=0$ as the equation which determines the background value of the preferred vector. In terms of the coefficients of \eqref{EFT_for_ST}, the conditions~$\barL_{\tilde{g}^{00}\tilde{g}^{00}} \neq 0$ and \eqref{solvable_con} are written as
\begin{align}
    M_2^4&\neq 0 \,, \quad 2\Mpl^2f+3\bar{M}_2^2+\bar{M}_3^2+\frac{3\bar{M}_1^6}{4M_2^4} \simeq 
    2\Mpl^2f+2\bar{M}_2^2+\frac{3\bar{M}_1^6}{4M_2^4}\neq 0\,,
    \label{solvable_conM}
\end{align}
where \eqref{subleading_scalingM} is assumed.

Now, we discuss consequences of the consistency conditions~\eqref{consistency1M}--\eqref{consistency4M}. Equations~\eqref{d_eq}, \eqref{lambda_eq}, and \eqref{conserve} lead to the relation
\begin{align}
    \dot{\Lambda}+3H (\dot{d}+\Mpl^2 \dot{f} H)=0
    \,,
\end{align}
which coincides with \eqref{consistency1M} with $\spatialR_{\rm BG}=0$ and $c=0$. Hence, \eqref{consistency1M} does not yield additional information. The consistency relation~\eqref{consistency2M} is written as
\begin{align}
    2M_2^4 \frac{\D}{\bar{N}\D t} \ln (-\tilde{g}^{00}_{\rm BG}) + 3\bar{M}_1^3 \dot{H} \simeq 0
    \,,
\end{align}
which determines $\frac{\D}{\bar{N}\D t} \ln (-\tilde{g}^{00}_{\rm BG})$. Then, \eqref{consistency3M} and \eqref{consistency4M} are
\begin{align}
    \bar{\rho}_{\rm m}+\bar{p}_{\rm m}+\dot{H}\left(2\Mpl^2f + 2\bar{M}_2^2 + \frac{3}{4} \frac{\bar{M}_1^6}{M_2^4} \right) &\simeq 0
    \,, \label{consistency_dotd}\\
    \Mpl^2 \dot{f}-\frac{3}{2}\dot{H} \frac{\bar{M}_1^3\mu_1^2}{M_2^4} &\simeq 0
    \,, \label{consistency_dotf}
\end{align}
where the background equations of motion are used. As expected, imposing condition~\eqref{solvable_conM}, we find de Sitter vacuum solution with $\dot{H}=0=\dot{f}$ in the absence of matter. The relations~\eqref{consistency_dotd} and \eqref{consistency_dotf} read the consistency relations between the background dynamics and the operators for the perturbations. For instance, if one considers the background with $\dot{f}=0$ and $\dot{H}\neq 0$, either $\delta \tilde{g}^{00}\delta \tilde{K}$ or $\delta \tilde{g}^{00} \delta \spatialR$ is not allowed at the leading order.

For comparison, we consider the scalar-tensor EFT for which the action is given by
\begin{align}
    S_{\rm ST}=\int \D^4x \sqrt{-g} \Biggl[
    &\frac{\Mpl^2}{2}f(t) \left( \spatialRST+K_{\mu\nu}K^{\mu\nu}-K^2 \right) - \Lambda(t)-c(t) g^{00}-d(t)K 
    \nn
    &+ \frac{1}{2} M_2^4(t) \left(\bar{N}^2\delta g^{00} \right)^2 + \cdots \Biggl]
    \,. \label{STEFT}
\end{align}
In the scalar-tensor EFT, the $c$-term and the $d$-term are not independent because the integration by parts of $d(t) K$ yields $n^{\mu}\partial_{\mu}d=\sqrt{-g^{00}} \partial_t d$. Whereas the $c$-term is retained in the conventional formulation of the EFT, we use the $d$-term to describe the background part of the EFT. The action of the scalar-tensor EFT becomes
\begin{align}
    S_{\rm ST}
    =\int \D^4x \sqrt{-g} \Biggl[
    &\frac{\Mpl^2}{2}f(t) \left( \spatialRST+K_{\mu\nu}K^{\mu\nu}-K^2 \right) - \Lambda_{\rm ST}(t)+\dot{d}_{\rm ST}(t)\sqrt{ 1-\bar{N}^2\delta g^{00} }  
    \nn
    &+ \frac{1}{2} M_{{\rm ST},2}^4(t) \left(\bar{N}^2\delta g^{00}\right)^2 + \cdots \Biggl]
    \,, 
    \label{STEFT2}
\end{align}
with
\begin{align}
\Lambda_{\rm ST}=\Lambda + \frac{c}{\bar{N}^2}\,, \quad 
\dot{d}_{\rm ST}=\dot{d}+\frac{2c}{\bar{N}^2}\,, \quad
M_{{\rm ST},2}^4=M_2^4+\frac{c}{2\bar{N}^2}
\,.
\label{redef_ST}
\end{align}
The background equations of motion are given by
\begin{align}
\dot{d}_{\rm ST}(t)&= - \bar{\rho}_{\rm m} - \bar{p}_{\rm m} - 2\Mpl^2(  f \dot{H} + \dot{f} H )
\,, \label{d_eq_ST} \\
\Lambda_{\rm ST}(t)&=-\bar{\rho}_{\rm m}+3 \Mpl^2  f H^2
\,, \label{lambda_eq_ST}
\end{align}
and we also have the matter conservation law~\eqref{conserve}. The above two equations have the same forms as \eqref{d_eq} and \eqref{lambda_eq}. In the general scalar-tensor theories without the shift symmetry, $\phi=t$ is a building block of the EFT and thus there is no consistency relations between the EFT coefficients. The $c$-function is completely absorbed into other coefficients according to \eqref{redef_ST}.

Note that the consistency relations~\eqref{consistency1M}--\eqref{consistency4M} in the vector-tensor EFT arise as a result of the absence of $t$ in the building blocks. Hence, the same relations hold for the shift-symmetric scalar-tensor theories~\cite{Nicolis:2011pv,Finelli:2018upr}. Equation~\eqref{consistency1M} and the background equations~\eqref{d_eq}, \eqref{lambda_eq}, and \eqref{conserve} give
\begin{align}
    \frac{\D}{\D t}\left( \frac{a^3 c(t)}{\bar{N}} \right) = 0
    \,, \label{scalar_eom}
\end{align}
which may be understood as the equation of motion of the scalar field in the shift-symmetric scalar-tensor theories. The shift symmetry ensures the existence of the constant of motion~$a^3c/\bar{N}$. In particular, $c=0$ is the attractor solution as the universe expands. This concludes that the background solution of the vector-tensor theories agrees with the attractor solution of the corresponding shift-symmetric scalar-tensor theories. The solution to \eqref{scalar_eom} is
\begin{align}
    c(t)=c_0 \frac{\bar{N}(t)}{a^3(t)}
    \,,
\end{align}
where $c_0$ is an integration constant. Then, the consistency relation~\eqref{consistency2M} becomes
\begin{align}
       -4M_{{\rm ST},2}^4 \frac{\dot{\bar{N}}}{\bar{N} }  + 3\bar{M}_1^3 \dot{H} +6 H \frac{c_0}{\bar{N}a^3}
    &\simeq 0
    \,, \label{consistency2ST}
\end{align}
while consistency relations~\eqref{consistency3M} and \eqref{consistency4M} turn out to be
\begin{align}
    \bar{\rho}_{\rm m}+\bar{p}_{\rm m}+\frac{2c_0}{\bar{N}a^3}+\dot{H}\left(2\Mpl^2f + 2\bar{M}_2^2 + \frac{3}{4} \frac{\bar{M}_1^6}{M_{{\rm ST},2}^4} \right) +H\frac{c_0}{\bar{N}a^3}\frac{3\bar{M}_1^3}{2M_{{\rm ST},2}^4}&\simeq 0
    \,, \label{consistency3ST}\\
    \Mpl^2 \dot{f}-\frac{\mu_1^2}{2M_{{\rm ST},2}^4} \left(3\bar{M}_1^3 \dot{H} + 6 H \frac{c_0}{\bar{N}a^3} \right) &\simeq 0
    \,, \label{consistency4ST}
\end{align}
where \eqref{d_eq_ST} and \eqref{consistency2ST} are used. Although $c=c_0\bar{N}/a^3 $ does not appear in the action~\eqref{STEFT2} explicitly, the $c$-function may affect the dynamics of perturbations through the consistency relations. Hereafter, we shall omit the subscript ST and use the unified notation between the scalar-tensor EFT and the vector-tensor EFT.

In summary, the background part of the action in both scalar-tensor and vector-tensor EFTs is given by
\begin{align}
    S=\int \D^4x \sqrt{-g} \Bigg[
    &\frac{\Mpl^2}{2}f(t) \left( \spatialRST+K_{\mu\nu}K^{\mu\nu}-K^2 \right) 
    + \bar{\rho}_{\rm m}-3\Mpl^2fH^2
    \nn
    &- \brb{\bar{\rho}_{\rm m}+\bar{p}_{\rm m}+2\Mpl^2(f\dot{H}+\dot{f}H)} \sqrt{ 1-\bar{N}^2\delta g^{00} } 
     + \cdots \Bigg] +S_{\rm m}
     \,,
    \label{LBG2}
\end{align}
after using the background equations of motion. At this stage, there is no clear distinction between the scalar-tensor theories and the vector-tensor theories. Note that the vector field has no dynamical mode in the homogeneous and isotropic configuration in vector-tensor theories and the dynamics of the background vector is determined by the constraint equation, $c=0$. On the other hand, the scalar field in scalar-tensor theories is dynamical and determined by a differential equation. This point is indeed important in vacuum spacetime: as we have seen, the vacuum solution in the vector-tensor theories has to be $H={\rm constant}$ while there can be a non-de Sitter solution in the scalar-tensor theories. However, in the context of EFT of dark energy, we additionally have a matter field which causes the time evolution of the universe. We may design the constraint equation of the vector-tensor theories to mimic the background dynamics of the scalar-tensor theories. Since we cannot replay the universe under different initial conditions, the EFT cannot distinguish whether the field is determined by a dynamical equation or a constraint. We need to discuss the perturbations to discriminate between the scalar-tensor EFT and the vector-tensor EFT.

\subsection{Linear scalar and tensor perturbations}
\label{sec:STpert}

Having obtained the background equations, we study the linear scalar and tensor perturbations in the vector-tensor EFT and discuss their characteristic properties. At the first order in perturbations, we have
\begin{align}
    \frac{\delta_1 \tilde{g}^{00}}{-\tilde{g}^{00}}=\bar{N}^2 \delta_1 g^{00} - \frac{2 \g}{1+\g \bar{A}_0} \delta A_0 \,,
\end{align}
where $\delta A_0$ is the perturbation of the vector field defined in \eqref{A_for_ST} and $\delta_1 \mathcal{Q}$ denotes the part of $\mathcal{Q}$ linear in the perturbations. Similarly, we will refer $\delta_2 \mathcal{Q}$ to be the part of $\mathcal{Q}$ which is quadratic in the perturbations. The EFT action does not contain the time derivative of $\delta A_0$. Therefore, the variable~$\delta A_0$ is an auxiliary variable and can be eliminated by using its equation of motion. Since the background part of the action~\eqref{LBG2} is independent of $\delta A_0$, we concentrate on $\mathcal{L}_{\rm DE}^{(2)}$, of which the quadratic part is given by
\begin{align}
     \delta_2 \mathcal{L}_{\rm DE}^{(2)}= \delta_2 \mathcal{L}_{\rm ST}^{(2)} + \delta_2 \mathcal{L}_{\delta A}^{(2)}\,,
\end{align}
where
\begin{align}
    \delta_2 \mathcal{L}_{\rm ST}^{(2)} &=\frac{1}{2}
      M_2^4 (\bar{N}^2\delta_1 g^{00})^2
    -  \frac{1}{2}{\bar M}_1^3 (\bar{N}^2\delta_1 g^{00}) \delta_1  K
-  \frac{1}{2}{\bar M}_2^2 (\delta_1 K)^2
\nn
& \quad
 - \frac{1}{2}{\bar M}_3^2 \delta_1 K^{\alpha}{}_{\beta} \delta_1 K^{\beta}{}_{\alpha} 
+\frac{1}{2}\mu_1^2 (\bar{N}^2\delta_1 g^{00})\delta_1 \! \spatialRST
+\frac{1}{2}\bar{M}_{4} \delta_1 K \delta_1 \! \spatialRST +\frac{1}{2}\lambda_{1} (\delta_1 \! \spatialRST)^2
\,, \\
    \delta_2 \mathcal{L}_{\delta A}^{(2)} &= \frac{1}{2}\left( \gf^2 M_2^4 +\frac{k^2}{a^2} \right) \delta \hat{A}_0^2 
    + \frac{1}{2} \gf \delta \hat{A}_0 \left[ -2 M_2^4 (\bar{N}^2\delta_1 g^{00}) +  \bar{M}_1^3 \delta_1 K - \mu_1^2 \delta_1 \! \spatialRST \right]\,,
    \label{L_effST}
\end{align}
in the momentum space with the comoving spatial momentum~$k=|k^i|$. Here, we have defined an effective gauge coupling~$\gf(t)$, of which mass dimension is $[\gf]=-1$, as
\begin{align}
    \gf(t) \equiv \frac{2 \g \bar{N}}{\sqrt{\gamma_1}(1+\g \bar{A}_0 ) }
    \,,
\end{align}
and the normalized field as $\delta \hat{A}_0 \equiv \sqrt{\gamma_1}\delta A_0/ \bar{N}$ provided $\gamma_1 >0$, respectively. As we will show in Sec.~\ref{sec:vector_pert}, $\gamma_1>0$ is required by the ghost-free condition of the vector perturbations; thus, $\gf$ is a real quantity and $\gf^2\geq 0$. Integrating out $\delta \hat{A}_0$, the quadratic Lagrangian of $\mathcal{L}_{\rm DE}^{(2)}$ becomes
\begin{align}
    \delta_2 \mathcal{L}_{\rm DE}^{(2)} &=\frac{1}{2}
      M_{\eff,2}^4(t,k) (\bar{N}^2\delta_1 g^{00})^2
    -  \frac{1}{2}\bar{M}_{\eff,1}^3(t,k) (\bar{N}^2\delta_1 g^{00}) \delta_1  K
    -  \frac{1}{2}\bar{M}_{\eff,2}^2(t,k) (\delta_1 K)^2
    \nn
    & \quad
    - \frac{1}{2}{\bar M}_3^2(t) \delta_1 K^{\alpha}{}_{\beta} \delta_1 K^{\beta}{}_{\alpha} 
    +\frac{1}{2}\mu_{\eff,1}^2(t,k) (\bar{N}^2\delta_1 g^{00}) \delta_1 \! \spatialRST
    \nn
    & \quad
    +\frac{1}{2}\bar{M}_{{\rm eff},4}(t,k) \delta_1 K  \delta_1 \! \spatialRST
    +\frac{1}{2}\lambda_{{\rm eff},1}(t,k) (\delta_1 \! \spatialRST)^2 \,,
    \label{Lquadratic_eff}
\end{align}
with the $k$-dependent coefficients,
\begin{align}
    M_{\eff,2}^4(t,k) &= \frac{k^2/a^2}{\gf^2 M_2^4+k^2/a^2} M_2^4 = [1-\mathcal{G}(t,k)] M_2^4(t)
    \,, \label{def_Meff2} \\
    \bar{M}_{\eff,1}^3(t,k)&= \frac{k^2/a^2}{\gf^2 M_2^4+k^2/a^2} \bar{M}_1^3 = [1-\mathcal{G}(t,k)] \bar{M}_1^3(t)
    \,, \\
    \bar{M}_{\eff,2}^2(t,k)&=  \bar{M}_2^2 + \frac{\gf^2 \bar{M}_1^6}{4(\gf^2 M_2^4 +k^2/a^2)} = \bar{M}_2^2(t) + \frac{1}{4}  \frac{\bar{M}_1^6(t)}{M_2^4(t)} \mathcal{G}(t,k)
    \,, \\
    \mu_{\eff,1}^2(t,k) &= \frac{k^2/a^2}{\gf^2 M_2^4+k^2/a^2} \mu_1^2 = [1-\mathcal{G}(t,k)] \mu_1^2(t)
    \,. \\
    \bar{M}_{{\rm eff},4}(t,k) &=\bar{M}_4 + \frac{\gf^2 \bar{M}_1^3 \mu_1^2}{2(\gf^2 M_2^4 +k^2/a^2)}= \bar{M}_4(t) +\frac{1}{2}\frac{\bar{M}_1^3(t) \mu_1^2(t)}{M_2^4(t)} \mathcal{G}(t,k)
    \,, \\
    \lambda_{{\rm eff},1}(t,k) &=\lambda_1-\frac{\gf^2 \mu_1^4}{4(\gf^2 M_2^4 +k^2/a^2)} = \lambda_1(t) -\frac{1}{4} \frac{\mu_1^4(t)}{M_2^4(t)} \mathcal{G}(t,k)
    \,,
    \label{def_lambda1}
\end{align}
where
\begin{align}\label{g-eff-def}
    \mathcal{G}(t,k) \equiv \frac{\gf^2 M_2^4}{\gf^2 M_2^4 +k^2/a^2} \,,
\end{align}
is a dimensionless function that controls the $k$-dependence of the coefficients. The function~$\mathcal{G}$ shows the following asymptotic behaviors:
\begin{align}
    \mathcal{G} \to 0 \quad {\rm as}\quad \frac{k^2/a^2}{\gf^2 M_2^4} \to \infty \qquad {\rm and} \qquad
    \mathcal{G} \to 1 \quad {\rm as}\quad  \frac{k^2/a^2}{\gf^2 M_2^4} \to 0\,.
\end{align}

After integrating out $\delta \hat{A}_0$, the Lagrangian~\eqref{Lquadratic_eff} takes the same form as the conventional EFT of inflation/dark energy, but the coefficients are $k$-dependent. An advantage of \eqref{Lquadratic_eff} is that we can universally describe the scalar-tensor EFT and the vector-tensor EFT by the same Lagrangian~\eqref{Lquadratic_eff} with the same background part~\eqref{LBG2}:
\begin{enumerate}
    \item The vector-tensor EFT corresponds to $\gf^2\neq 0$. We have to impose the consistency relations~\eqref{consistency_dotd} and \eqref{consistency_dotf} between the coefficients for perturbations and the background dynamics. Note, however, that the Lagrangian~\eqref{Lquadratic_eff} is available only for irrotational solutions. The vector perturbations have to be separately discussed (see Sec.~\ref{sec:vector_pert}). 
    \item The general scalar-tensor EFT is obtained by setting $\gf^2=0$ without imposing the consistency relations between the EFT coefficients. The $k$-dependence of the coefficients disappear and one recovers the conventional EFT of inflation/dark energy.
    \item The shift-symmetric scalar-tensor EFT corresponds to $\gf^2=0$ under the consistency relations~\eqref{consistency3ST} and \eqref{consistency4ST} where $M_{{\rm ST},2}^4(t)$ is understood as $M_2^4(t)$ in the current notation. The constant parameter $c_0$ represents the deviation from the attractor solution in an expanding universe.
\end{enumerate}
Hence, three different classes of EFTs are distinguished by the function~$\gf^2(t)$ and the consistency relations which can be used to describe inflation/dark energy models. Note that the consistency relations in the vector-tensor theories and those in the shift-symmetric scalar-tensor theories are generically different due to the parameter~$c_0$. In the case of an expanding universe, when we restrict our consideration to the attractor solution of the shift-symmetric theories, i.e.~$c_0=0$, the consistency relations agree with each other. Then, the function~$\gf^2(t)$ is the only function to discriminate the vector-tensor theories from the shift-symmetric scalar-tensor theories.

It would be useful to mention general features of perturbations before performing the detailed analysis. First of all, the operators~$\delta g^{00}$, $\delta K$, and $\delta_1 \! \spatialRST$ are unperturbed by the tensor perturbations. The EFT coefficient~$\bar{M}_3^2$ is the only relevant parameter for the modification of the tensor perturbations which has no $k$-dependence. Hence, there is no essential difference from the scalar-tensor EFT in the linear tensor perturbations. We need to study the scalar perturbations (and the vector perturbations) to see differences between the scalar-tensor theories and the vector-tensor theories.

We have assumed \eqref{scalingM} and \eqref{subleading_scalingM}, that is, $\bar{M}_2^2+\bar{M}_3^2$, $\bar{M}_4$, and $\lambda_1$ are supposed to be suppressed by the cutoff scale~$\cutoff$. The functions with ``eff'' read
\begin{align}
    \bar{M}_{{\rm eff},2}^2+\bar{M}_{{\rm eff},3}^2  &\sim \Mpl^2\left( \frac{H^2}{\cutoff^2} + \mathcal{G} \right)
    \,, ~
    \bar{M}_{{\rm eff},4} \sim \Mpl^2H^{-1} \left( \frac{H^2}{\cutoff^2} + \mathcal{G} \right)
    \,, ~
    \lambda_{{\rm eff},1} \sim \Mpl^2H^{-2} \left( \frac{H^2}{\cutoff^2} + \mathcal{G} \right)
    \,,
    \label{high_eff}
\end{align}
which do not vanish even if we take the limit $\cutoff \to \infty$. The ``higher-derivative'' operators should be kept in the vector-tensor EFT.
However, we should carefully look at the operators corresponding to \eqref{high_eff}. For instance, assuming $H^2\lesssim \gf^2M_2^4$, let us consider $(\delta_1 \! \spatialRST)^2$ in the scales~$\gf^2 M_2^4 \ll k^2/a^2 \ll \cutoff^2$. The term suppressed by $\cutoff$ can be ignored and $\mathcal{G}$ is approximated as $\mathcal{G}\approx a^2\gf^2M_2^4/k^2\propto k^{-2}$. We then obtain
\begin{align}
    \lambda_{{\rm eff},1} (\delta_1 \! \spatialRST)^2 \propto k^2\,,
\end{align}
where we have used $\delta_1\! \spatialRST =\mathcal{O}(k^2)$. Therefore, in this regime, the operator~$(\delta_1 \! \spatialRST)^2$ is not a higher-derivative term due to the non-local $k$-dependence of the coefficient. On the other hand, the $k$-dependence of the other coefficients, say $M_{{\rm eff},2}^4$, may be ignored in the regime~$\gf^2 M_2^4 \ll k^2/a^2 \ll \cutoff^2$ because $M_{{\rm eff},2}^4=M_2^4+\mathcal{O}(k^{-2})$. This observation implies that the deviation from the scalar-tensor EFT on small scales may be caused by the operators corresponding to \eqref{high_eff} and the non-local $k$-dependence of $\mathcal{G}$.

The non-local $k$-dependence also plays an important role on large scales, $k^2/a^2 \ll \gf^2 M_2^4$, for which we have
\begin{align}
    M_{\rm eff,2}^4,~ \bar{M}_{\rm eff,1}^3, ~ \mu_{\rm eff,1}^2 \propto k^2 \to 0 \,, \qquad
    {\rm as} \qquad \frac{k^2/a^2}{\gf^2 M_2^4} \to 0 \,.
    \label{asympt_Meff}
\end{align}
The operators associated with $\delta g^{00}$ vanish in the limit~$k\to 0$. This property can be understood as follows: the variable $\delta A_0$ appears only through $\delta \tilde{g}^{00}=\mathcal{O}(\delta A_0)$ or $F_{\mu\nu}=\mathcal{O}(\partial_i \delta{A}_0)$. At sufficiently large scales, we can ignore contributions from $F_{\mu\nu}=\mathcal{O}(\partial_i \delta{A}_0)$ in comparison with the ones from $\delta \tilde{g}^{00}$. We then use $\delta \tilde{g}^{00}$ as an independent variable instead of $\delta A_0$. The invertibility of the transformation is guaranteed by the non-vanishing norm of the preferred vector. Since $F_{\mu\nu}$ can be ignored, the variation with respect to $\delta \tilde{g}^{00}$ determines $\delta \tilde{g}^{00}$ as a function of other building blocks. As a result, the action on large scales does not explicitly depend on $\delta g^{00}$, in agreement with the behavior of coefficients in \eqref{asympt_Meff}.

The operators involving $\delta g^{00}$ yield the kinetic term of the scalar mode. Hence, there is no dynamical degree of freedom exactly at $k=0$, consistently with the fact that the vector field is non-dynamical in the homogeneous and isotropic ansatz. One needs to retain the finite $k$-dependence to study the dynamics of the scalar perturbations in vector-tensor EFT. As a simple but non-trivial example, we consider the gauged ghost condensate only with $M_2^4$ and $\bar{M}_2^2$:
\begin{align}
    \delta_2 \mathcal{L}_{\rm DE}^{(2)} &=\frac{1}{2}
      M_{\eff,2}^4(t,k) (\bar{N}^2\delta_1 g^{00})^2
-  \frac{1}{2}\bar{M}_{\eff,2}^2(t,k) (\delta_1 K)^2
+\cdots
\nn
&\sim \frac{k^2/a^2}{\gf^2 M_2^4+k^2/a^2}M_2^4 \dot{\pi}^2 - \bar{M}_2^2 \frac{k^4}{a^4} \pi^2
\nn
&\sim \frac{k^2/a^2}{\gf^2 M_2^4+k^2/a^2}\left[ M_2^4 \dot{\pi}^2 - \bar{M}_2^2 \left( \gf^2 M_2^4 + \frac{k^2}{a^2} \right) \frac{k^2}{a^2} \pi^2 \right]\,,
\end{align}
where $\pi$ is the scalar mode and we have used $\bar{N}^2\delta_1 g^{00} \sim \dot{\pi}$ and $\delta_1 K \sim \frac{k^2}{a^2} \pi$. Compared with the ghost condensate where $\gf^2=0$, the dispersion relation acquires the $k^2$~term due to the non-local $k$-dependence of the kinetic term. The large scale behavior of the perturbations is changed by the operators associated with $\delta g^{00}$.

\subsection{Phenomenological predictions}
\label{sec:pheno_prediction}

\subsubsection{Phenomenological parameterization}
We now discuss the phenomenological implications of the vector-tensor EFT in the context of dark energy. An alternative parameterization, dubbed the $\alpha$-basis, has been commonly used for the phenomenological study of dark energy~\cite{Bellini:2014fua,Gleyzes:2014qga,Gleyzes:2014rba,Frusciante:2016xoj}. We define the following functions:
\begin{align}
    \begin{split}
    &\tilde{\alpha}_B(t,k) \equiv -\frac{\bar{M}_{{\rm eff},1}^3}{2H M^2}
    \,, \quad
    \alpha_T(t) \equiv \frac{\bar{M}_{3}^2}{M^2}
    \,, \quad
    \tilde{\alpha}_K(t,k)\equiv  \frac{4M_{{\rm eff},2}^4}{H^2 M^2}
    \,, \\
    &\tilde{\alpha}_H(t,k) \equiv \frac{2\mu_{{\rm eff},1}^2+\bar{M}_3^2}{M^2}
    \,, \quad
    \tilde{\alpha}_B^{\rm GLPV}(t,k) \equiv \frac{\bar{M}_3^2+\bar{M}_{{\rm eff},2}^2}{M^2}
    \,, 
    \\
    &\tilde{\alpha}_M^{\rm GC}(t,k) \equiv \frac{\lambda_{{\rm eff},1} H^2}{M^2}
    \,, \quad
    \tilde{\alpha}_B^{\rm GC}(t,k) \equiv \frac{\bar{M}_{{\rm eff},4} H}{M^2}
    \,,
    \end{split}
    \label{def_alpha}
\end{align}
where
\begin{align}
    M^2(t) \equiv \Mpl^2 f - \bar{M}_3^2
    \,, \label{def_M}
\end{align}
is the effective Planck mass for the tensor perturbations.\footnote{The definitions of $\tilde{\alpha}_B$ and $\tilde{\alpha}_K$ are apparently different from those in~\cite{Bellini:2014fua,Gleyzes:2014qga,Gleyzes:2014rba,Frusciante:2016xoj}, but they are equivalent in the scalar-tensor limit (we follow the convention of \cite{Gleyzes:2014rba,Gleyzes:2014qga}). Recall that the background part of our EFT action~\eqref{LBG2} is different from the conventional parameterization of the scalar-tensor EFT.} By means of the functions defined by \eqref{def_alpha}, the quadratic action for the scalar and tensor perturbations in the momentum space is given by
\begin{align}
    \delta_2 S = \int &\frac{\D t \D^3k}{(2\pi)^3} \bar{N}a^3  \frac{M^2}{2}
    \nn
    \times \Biggl[ & 
    (1+\tilde{\alpha}_H) \frac{\delta N}{\bar{N}} \delta_1 \! \spatialRST + 4 H \tilde{\alpha}_B \frac{\delta N}{\bar{N}} \delta_1 K
    +\delta_1 K^{\alpha}{}_{\beta} \delta_1 K^{\beta}{}_{\alpha}
    -(1+\tilde{\alpha}_B^{\rm GLPV}) (\delta_1 K)^2 
    \nn
    &
    + \tilde{\alpha}_K H^2 \left(\frac{\delta N}{\bar{N}}\right)^2
    +(1+\alpha_T) \delta_2\bigg( \spatialRST \frac{\sqrt{h}}{a^3}\, \bigg)  + \frac{\tilde{\alpha}_M^{\rm GC}}{H^2} (\delta_1 \! \spatialRST)^2
    + \frac{\tilde{\alpha}_B^{\rm GC} }{H} \delta_1 K  \delta_1 \! \spatialRST \Biggl] 
    + \delta_2 S_{\rm m} + \cdots \,,
    \label{LEFT_alpha}
\end{align}
where $\delta N \equiv N-\bar{N}$ is the perturbation of the lapse and the ellipsis represents contributions from the higher-order operators which are suppressed by the cutoff~$\cutoff$. It is convenient to introduce the parameter representing the evolution rate of the effective Planck mass,
\begin{align}
    \alpha_M(t) \equiv \frac{1}{H} \frac{\D \ln M^2(t)}{\bar{N} \D t}
    \,.
    \label{def_alphaM}
\end{align}
We also introduce the functions without the tilde which are defined through the relations analogous to \eqref{def_alpha} by the use of the EFT coefficients without ``eff'', e.g.~$\alpha_B(t) \equiv -\bar{M}_{1}^3/(2H M^2)$. The functions with the tilde depend on $k$ while the functions without the tilde do not. The $\alpha$-functions are dimensionless and characterize the various effects of modification of gravity. In terms of the $\alpha$-functions, the consistency relations~\eqref{consistency_dotd} and \eqref{consistency_dotf} read
\begin{align}
    \bar{\rho}_{\rm m}+\bar{p}_{\rm m} + 2M^2 \dot{H} \frac{\alpha_K + 6\alpha_B^2}{\alpha_K} &\simeq 0
    \,, \label{consistency1_alpha} \\
    \Mpl^2 \dot{f} + 6M^2 \frac{\dot{H}}{H} \frac{ \alpha_B (\alpha_H-\alpha_T)}{\alpha_K} &\simeq 0
    \,, \label{consistency2_alpha}
\end{align}
with
\begin{align}
\Mpl^2 \dot{f}=M^2[ H\alpha_M(1+\alpha_T)+\dot{\alpha}_T ]
\,.
\end{align}

As discussed in subsection~\ref{sec:STpert}, the signature of the vector-tensor EFT is embedded in the non-local $k$-dependence of the EFT coefficients and the behavior of the system changes qualitatively at the critical scale~$\gf^2 M_2^4$. The dependence on $k$ are encoded in $\mathcal{G}(t,k)$ which is defined in \eqref{g-eff-def}. In terms of the $\alpha$-functions, we find 
\begin{align}\label{g-eff}
    \mathcal{G}(t,k)=\frac{\alpha_V \alpha_K}{\alpha_V \alpha_K + k^2/(a^2H^2)}
    \,,
\end{align}
where we have defined a new dimensionless $\alpha$-function
\begin{align}
    \alpha_V \equiv \frac{1}{4}\gf^2 M^2 
    \,.
    \label{def_alphaV}
\end{align}
Then, the $\alpha$-functions with the tilde can be written in terms of the $\alpha$-functions without the tilde as follows
\begin{align}
    \begin{split}
    \tilde{\alpha}_B(t,k)& = [1-\mathcal{G}(t,k)]\alpha_B(t)\,, \\
    \tilde{\alpha}_K(t,k) &= [1-\mathcal{G}(t,k)]\alpha_K(t) \,, \\
    \tilde{\alpha}_H(t,k) &= [1-\mathcal{G}(t,k)]\alpha_H(t) + \mathcal{G}(t,k) \alpha_T(t)
    \,, \\
    \tilde{\alpha}_B^{\rm GLPV}(t,k)  &= \alpha_B^{\rm GLPV}(t) + 4 \mathcal{G}(t,k) \frac{\alpha_B^2(t)}{\alpha_K(t)}
    \,, \\
    \tilde{\alpha}_M^{\rm GC}(t,k) &= \alpha_M^{\rm GC}(t)-\frac{1}{4}\mathcal{G}(t,k) \frac{[\alpha_H(t)-\alpha_T(t)]^2}{\alpha_K(t)}
    \,, \\
    \tilde{\alpha}_B^{\rm GC}(t,k) &= \alpha_B^{\rm GC}(t)-2 \mathcal{G}(t,k) \frac{\alpha_B(t)[\alpha_H(t)-\alpha_T(t)]}{\alpha_K(t)}
    \,.
    \end{split}
    \label{alpha_relation}
\end{align}
The relations~\eqref{alpha_relation} provide the mapping between the vector-tensor EFT and the scalar-tensor EFT in the $\alpha$-basis. The scalar-tensor EFT is recovered for $\mathcal{G}(t,k)=0$ or $\alpha_V =0$. The ghost-free conditions conclude $\alpha_V>0$ as we will see, so we need to consider only the positive region for the phenomenological purpose.  Then, the general theories of dark energy, including not only scalar-tensor theories but also vector-tensor theories, are characterized by the $4+3+1$ time-dependent functions,
\begin{align}
    \begin{split}
    \alpha_K(t)\,, \quad 
    \alpha_B(t)\,, \quad
    \alpha_T(t)\,, \quad
    \alpha_H(t)\,, \\
    \alpha_B^{\rm GLPV}(t)\,, \quad \alpha_M^{\rm GC}(t)\,, \quad \alpha_B^{\rm GC}(t)\,, \\
    \alpha_V(t) \,,
    \end{split}
\end{align}
in addition to the background EFT coefficients. In particular, the generalized Proca theory corresponds to
\begin{align}
    \alpha_K,\alpha_B,\alpha_T,\alpha_V \neq 0 
    \,, \quad
    \alpha_H=\alpha_B^{\rm GLPV}=\alpha_M^{\rm GC}=\alpha_B^{\rm GC}=0
    \,.
\end{align}
The function~$\alpha_H$ represents the deviation from the Horndeski theory in the scalar-tensor case~$(\alpha_V=0)$ and from the generalized Proca theory~$(\alpha_V \neq 0)$ in the vector-tensor case. 
The functions~$\alpha_B^{\rm GLPV}$, $\alpha_M^{\rm GC}$, $\alpha_B^{\rm GC}\propto \cutoff^{-2}$ are those for the (gauged) ghost condensate which can be discarded at the leading order of the derivative expansion when we are interested in backgrounds away from the (gauged) ghost condensate.

\subsubsection{Stability conditions}
\label{sec:stability}
The stability conditions for the tensor perturbations can be immediately found from the action~\eqref{LEFT_alpha} as
\begin{align}
    M^2>0\,, \quad 1+\alpha_T >0
    \,,
\end{align}
while the vector perturbations will be explored in the next section. 

We thus concentrate on the scalar perturbations in which the metric perturbations are given by
\begin{align}
    \delta g_{00}=-2\bar{N}^2 \alpha\,, \quad
    \delta g_{0i}=\bar{N}a \partial_i \beta\,, \quad
    \delta g_{ij}=2a^2 \zeta \delta_{ij}
    \,.
\end{align}
Here, the spatial diffs are used to diagonalize the spatial metric and the combined time diffs and $U(1)$ has been used to eliminate the spatial components of the vector field~$A_{\mu}$.

As for the matter action, we consider the k-essence
\begin{align}
    S_{\rm m}=\int \D^4x \sqrt{-g}P(X_{\sigma})\,, \quad X_{\sigma}=-\frac{1}{2}(\partial \sigma)^2
    \,, \label{k-essence}
\end{align}
to represent the irrotational perfect fluid. 
The scalar field~$\sigma$ is decomposed into
\begin{align}
    \sigma(t,x^i) =\bar{\sigma}(t)+\delta\sigma(t,x^i)\,, \quad
    \delta\sigma(t,x^i)\equiv \frac{\dot{\bar{\sigma}}}{H} \delta \hat{\sigma}(t,x^i)
    \,,
\end{align}
where the perturbation~$\delta \hat{\sigma}$ is normalized to be dimensionless. We recall that a dot represents the derivative with respect to the cosmic time, $\dot{\bar{\sigma}}=\D\bar{\sigma}/(\bar{N}\D t)$. The background energy density, pressure, and the squared sound speed are given by
\begin{align}
    \bar{\rho}_{\rm m} = 2\bar{X}\bar{P}_X-\bar{P}\,, \quad \bar{p}_{\rm m} = \bar{P}\,, \quad
    c_{\rm m}^2 = \frac{\bar{P}_X}{\bar{P}_X + 2\bar{X} \bar{P}_{XX}} \,,
\end{align}
respectively, where the quantities with the bar are the background quantities and the subscript~$X$ represents the derivative with respect to $X_{\sigma}$, e.g.~$P_X=\D P/\D X_{\sigma}$.

The quadratic action is a functional of four variables, $\alpha$, $\beta$, $\zeta$, and $\delta \hat{\sigma}$. The perturbations of the lapse and the shift, $\alpha$ and $\beta$, are non-dynamical and can be integrated out. As a result, the quadratic action takes the form
\begin{align}
    \delta_2 S = \int \frac{\D t \D^3k}{(2\pi)^3}\bar{N}a^3\left[ \frac{1}{2}\dot{\xi}^t {\bf K}_S\dot{\xi} + \xi^t {\bf M}_S \dot{\xi} - \frac{1}{2}\xi^t {\bf V}_S \xi
    \right]\,, \quad 
    \xi = 
    \begin{pmatrix}
    \zeta \\ \delta \hat{\sigma}
    \end{pmatrix}
    \,,
\end{align}
where ${\bf K}_S$ and ${\bf V}_S$ are symmetric matrices while ${\bf M}_S$ is an antisymmetric matrix. The absence of ghost and gradient instabilities at small scales requires the boundedness of the Hamiltonian in the high-momentum limit which is guaranteed by the positive definiteness of the matrices~${\bf K}_S$ and ${\bf V}_S$ in the high-momentum limit. In order to justify the derivative expansion, we consider the high-momentum modes satisfying
\begin{align}
    \frac{k^2}{a^2} \gg H^2 \,, \qquad \frac{k^2}{a^2} \gg \gf^2 M_2^4 = \alpha_V \alpha_K H^2
    \,, \qquad \mbox{with} \qquad  \frac{k^2}{a^2}\ll \cutoff^2\,.
\end{align}
The asymptotic forms of the matrices are then given by
\begin{align}
    {\bf K}_S = {\bf K}_0 + \mathcal{O}(k^{-2})\,, \quad
    {\bf M}_S = \frac{k^2}{a^2}  {\bf M}_2 +  \mathcal{O}(k^0)\,, \quad
    {\bf V}_S =  \frac{k^4}{a^4} {\bf V}_4 +  \frac{k^2}{a^2} {\bf V}_2 +  \mathcal{O}(k^0) \,.
\end{align}
The high-energy limit dispersion relation is given by roots of
\begin{align}
    {\rm det}\left[ \omega^2{\bf K}_0+i\omega \frac{k^2}{a^2}({\bf M}_2-{\bf M}_2^t)-\frac{k^4}{a^4}{\bf V}_4 -\frac{k^2}{a^2}{\bf V}_2\right] =0
    \,.
\end{align}
The higher-derivative terms, ${\bf M}_2$ and ${\bf V}_4$, are generated by $\alpha_B^{\rm GLPV}$, $\alpha_M^{\rm GC}$, and $\alpha_B^{\rm GC}$, meaning that
\begin{align}
    {\bf M}_2,~{\bf V}_4 \propto \cutoff^{-2}\,,
\end{align}
under the assumption \eqref{subleading_scalingM}. Hence, we may ignore the higher-derivative corrections as for theories with ${\cal O}({\bf V}_2)/{\cal O}({\bf K}_0)=\mathcal{O}(1)$. We first consider the theories with ${\cal O}({\bf V}_2)/{\cal O}({\bf K}_0)=\mathcal{O}(1)$ for which we can practically set $\alpha_B^{\rm GLPV}=\alpha_M^{\rm GC}=\alpha_B^{\rm GC}=0$. Then, we find the relativistic dispersion relations for the dark energy field and the matter field. The stability conditions, ${\bf K}_0>0$ and ${\bf V}_2>0$, yield
\begin{align}
    K_S>0 \,, \quad V_S>0
    \,,
\end{align}
on top of the stability conditions of the matter field~$\bar{\rho}_{\rm m}+\bar{p}_{\rm m}>0$ and $c_{\rm m}^2>0$, where
\begin{align}
    K_S &\equiv {\rm det}\left[ {\bf K}_0\right] = \alpha_K + 6\alpha_B^2
    \,, \\
    V_S &\equiv {\rm det}\left[ {\bf V}_2\right] = 4\alpha_V\mathcal{A}^2
    +2(1+\alpha_B)\mathcal{A}
    \nn
    &\quad -2(1+\alpha_H)^2\left[ \frac{\bar{\rho}_{\rm m}+\bar{p}_{\rm m}}{2M^2H^2} + \frac{1+\alpha_B}{1+\alpha_H}\left(\frac{\dot{H}}{H^2}-\alpha_M \right)+\frac{1}{H}\frac{\D}{\bar{N}\D t} \left(\frac{1+\alpha_B}{1+\alpha_H} \right) \right]
    \,, \label{def_VS}
\end{align}
with
\begin{align}
    \mathcal{A}\equiv \alpha_H-\alpha_T-\alpha_B(1+\alpha_T)
    \,. \label{def_calA}
\end{align}
The function~$\alpha_V$ enters the stability condition~$V_S>0$ and change the sound speed of dark energy. The dispersion relations are roots of
\begin{align}
   \left(\omega^2 - c_{\rm m}^2 \frac{k^2}{a^2} \right)\left( K_S \omega^2 - V_S \frac{k^2}{a^2}\right)
   -\alpha_H^2 \frac{\bar{\rho}_{\rm m}+\bar{p}_{\rm m}}{M^2H^2}\frac{\omega^2 k^2}{a^2} =0
   \,.
\end{align}
In particular, when $\alpha_H=0$, which is the case of generalized Proca theories, the sound speed of dark energy is simply given by $c_S^2=V_S/K_S$.

In the context of degenerate scalar-tensor theories, it is known that the perturbations about the stealth solutions generically do not yield the $k^2$~term, which signals strong coupling~\cite{Motohashi:2019ymr,Takahashi:2021bml}.
In this case, one needs to take into account the scordatura effect, i.e.~weak and controlled violation of the degeneracy condition~\cite{Motohashi:2019ymr}.
The scordatura introduces the $k^4$~term in the dispersion relation as in ghost condensate~\cite{Arkani-Hamed:2003pdi,Arkani-Hamed:2003juy}, which can cure the strong coupling issue while keeping the ghost-free nature below the EFT cutoff scale.
In our EFT framework, there are also theories like ghost condensate or scordatura \cite{Arkani-Hamed:2003juy,Arkani-Hamed:2003pdi,Mukohyama:2006be,Motohashi:2019ymr,Gorji:2020bfl,Gorji:2021isn} in which ${\cal O}({\bf V}_2)/{\cal O}({\bf K}_0)\ll{\cal O}(1)$. In that case, the higher-order operators labeled by $\alpha_B^{\rm GLPV}$, $\alpha_M^{\rm GC}$, and $\alpha_B^{\rm GC}$ become dominant and, therefore, we have to take into account their roles. To be more concrete, let us consider the de Sitter limit for which the second line of \eqref{def_VS} vanishes due to the time translation invariance. Imposing the condition~$\mathcal{A}=\alpha_H-\alpha_T-\alpha_B(1+\alpha_T)=0$, the sound speed of the dark energy field vanishes at the leading order in the derivative expansion. Then, the originally sub-leading terms~${\bf M}_2$ and ${\bf V}_4$ become the dominant terms. The dispersion relation of dark energy in the (gauged) ghost condensate phase is given by
\begin{align}
\omega^2=\frac{\alpha_B^{\rm GLPV}(1+\alpha_H)^2+4\alpha_B^{\rm GC}(1+\alpha_H)(1+\alpha_B)-16\alpha_M^{\rm GC}(1+\alpha_B)^2}{H^2(\alpha_K+6\alpha_B^2)} \frac{k^4}{a^4}
+\mathcal{O}\brp{\frac{k^2 H^2}{\cutoff^2}}
\,,
\end{align}
where we have used $\alpha_B^{\rm GLPV}, \alpha_M^{\rm GC}, \alpha_B^{\rm GC} = \mathcal{O}(H^2/\cutoff^2) \ll 1$. The stability condition is guaranteed by $K_S>0$ and the positivity of the $k^4$ coefficient in the above dispersion relation.

\subsubsection{Effective gravitational coupling and slip parameter}

In order to look for the observational consequences of the vector-tensor EFT of dark energy, we compute the quadratic action for the scalar perturbations under the quasi-static approximation. We define the Newtonian gauge variables:\footnote{Since we will focus on the sub-horizon scales, the gauge choice of the variables might be inessential. Nevertheless, it would be convenient to introduce the Newtonian gauge variables in order to compare with other studies. Here, we follow the convention used in~\cite{Planck:2018vyg}.}
\begin{align}
    \Psi \equiv \alpha + a \dot{\beta} + aH\beta \,, \quad
    \Phi \equiv -\zeta - a H \beta
    \,,
\end{align}
and
\begin{align}
    \hat{\pi} \equiv a H \beta
    \,.
\end{align}
In this notation, the spacetime metric in the Newtonian gauge is written as
\begin{align}
    \D s^2=-\bar{N}^2 (1+2\Psi)\D t^2+a^2(t) (1-2\Phi) \delta_{ij} \D x^i \D x^j
    \,.
\end{align}
The dimensionless variable $\hat{\pi}$ corresponds to the spatial component of the vector field~$A_{\mu}$ in the Newtonian gauge, representing the scalar mode of dark energy.

Regarding the matter variable, we define the comoving density contrast~$\Delta$ as
    \be
    \Delta \equiv \frac{\delta \rho_{\rm m}}{\bar{\rho}_{\rm m}}+3H\frac{\bar{\rho}_{\rm m}+\bar{p}_{\rm m}}{\bar{\rho}_{\rm m}}  \frac{\delta\sigma}{\dot{\bar{\sigma}}}
    \,,
    \ee
where, for a k-essence matter described by \eqref{k-essence}, $\delta \rho_{\rm m}$ is given by
    \be
    \delta \rho_{\rm m}=\frac{\bar{\rho}_{\rm m}+\bar{p}_{\rm m}}{c_{\rm m}^2}\brp{\frac{\dot{\delta\sigma}}{\dot{\bar{\sigma}}}-\alpha}\,.
    \ee
In what follows, we consider a dust fluid with $\bar{p}_{\rm m}=0$ and $c_{\rm m}=0$.
The matter Lagrangian can be derived from the k-essence Lagrangian by changing the variable from $\delta\sigma$ to $\Delta$ and then taking the dust limit~(see e.g.~Appendix~A of \cite{DeFelice:2015moy}).

The quadratic action for the scalar perturbations is given by
\begin{align}
    \delta_2 S\simeq \int \frac{\D t \D^3k}{(2\pi)^3} \bar{N}a^3 \left[\mathcal{L}_{\rm DE}^{(QS)}+\mathcal{L}_{\rm m}^{(QS)} \right]\,,
    \label{LSS}
\end{align}
with
\begin{align}
    \mathcal{L}_{\rm DE}^{(QS)}=
    \frac{M^2}{2} \frac{k^2}{a^2}\Biggl\{
    &\frac{k^2}{a^2H^2}\left[ 4(8\tilde{\alpha}_M^{\rm GC}-\tilde{\alpha}_B^{\rm GC})\Phi \hat{\pi} - (\tilde{\alpha}_B^{\rm GLPV} - 16\tilde{\alpha}_M^{\rm GC} + 4\tilde{\alpha}_B^{\rm GC}) \hat{\pi}^2 + 16 \tilde{\alpha}_M^{\rm GC} \Phi^2\right]
    \nn
    &
    -\tilde{c}_1 \Psi \hat{\pi} - \tilde{c}_2 \Phi \hat{\pi} - \tilde{c}_3\hat{\pi}^2 
    -\tilde{c}_4 \Phi \Psi - \tilde{c}_5 \Phi^2
    -\tilde{c}_6 \frac{\dot{\Phi}}{H}\hat{\pi}
    \Biggl\}\,,
    \label{LDE_SS}
\end{align}
under the quasi-static approximation, where we have introduced
\begin{align}
    \tilde{c}_1(t,k) &\equiv 2(2\tilde{\alpha}_H-2\tilde{\alpha}_B - 3\tilde{\alpha}_B^{\rm GLPV} - 6\tilde{\alpha}_B^{\rm GC})
    \,, \label{def_c1} \\
    \tilde{c}_2(t,k) &\equiv 4\left[ \tilde{\alpha}_H(1+\alpha_M)+ \alpha_M - \alpha_T +\frac{\dot{\tilde{\alpha}}_H}{H}  - 3\tilde{\alpha}_B^{\rm GC} \frac{\dot{H}}{H^2} \right]
    \,, \\
    \tilde{c}_3(t,k) & \equiv
    -2\Biggl[ \left( 1+\alpha_M+ \frac{\dot{H}}{H^2} \right)(\tilde{\alpha}_B-\tilde{\alpha}_H) + \frac{\dot{\tilde{\alpha}}_B - \dot{\tilde{\alpha}}_H }{H} + \frac{\bar{\rho}_{\rm m}}{2M^2H^2}
    \nn
    &\qquad \quad +(1+ 3\tilde{\alpha}_B^{\rm GLPV}+ 6 \tilde{\alpha}_B^{\rm GC}) \frac{\dot{H}}{H^2} + \alpha_T -\alpha_M
    \Biggl]
    \,, \\
    \tilde{c}_4(t,k) &\equiv 4(1+\tilde{\alpha}_H-3\tilde{\alpha}_B^{\rm GC})
    \,, \\
    \tilde{c}_5(t,k) &\equiv - 2\left[1+\alpha_T -3\tilde{\alpha}_B^{\rm GC}\left(1+\alpha_M - \frac{\dot{H}}{H^2} \right) - \frac{3\dot{\tilde{\alpha}}_B^{\rm GC}}{H} \right]
    \,, \\
    \tilde{c}_6(t,k) &\equiv 2(2\tilde{\alpha}_H-3\tilde{\alpha}_B^{\rm GLPV}-6\tilde{\alpha}_B^{\rm GC})
    \,. \label{def_c6}
\end{align}
As for the matter field, the quadratic Lagrangian for the comoving density contrast of the dust fluid is given by
\begin{align}
    \mathcal{L}_{\rm m}^{(QS)} = -\bar{\rho}_{\rm m} \Psi \Delta + \frac{1}{2}\frac{a^2\bar{\rho}_{\rm m} \dot{\Delta}^2}{k^2}
    \,,
\end{align}
at the sub-horizon scales, which yields the standard equation of motion of the density contrast,
\begin{align}
    \ddot{\Delta}+2H\dot{\Delta}+\frac{k^2}{a^2}\Psi=0
    \,.
\end{align}

We recall that the $\alpha$-functions with the tilde are related to the $\alpha$-functions without the tilde via \eqref{alpha_relation} through the $k$-dependence encoded in $\mathcal{G}(t,k)$. Hence, if the phenomenological functions~$\alpha_K$ and $\alpha_V$ are of order unity, the function~$\mathcal{G}$, defined in \eqref{g-eff} is approximated by
\begin{align}
    \mathcal{G}\simeq \alpha_V \alpha_K \frac{a^2H^2}{k^2}
    \,,
\end{align}
at the scales~$k^2/a^2H^2\gg1$. The quasi-static limit is, however, $c_S^2k^2/a^2H^2\gg1$ and the condition~$k^2/a^2H^2\gg1$ does hold for $c_S\ll1$ [or more generally for ${\cal O}({\bf V}_2)/{\cal O}({\bf K}_0)\ll{\cal O}(1)$]. In this case, the operators labeled by $\alpha_B^{\rm GLPV}, \alpha_M^{\rm GC}, \alpha_B^{\rm GC}$ make the quasi-static limit well-defined~\cite{Gorji:2020bfl}. Here for simplicity we consider $c_S={\cal O}(1)$ (or ${\cal O}({\bf V}_2)/{\cal O}({\bf K}_0)={\cal O}(1)$) so that the quasi-static limit coincides with $k^2/a^2H^2\gg1$. Then, the first line of \eqref{LDE_SS} is
\begin{align}
    &\frac{k^2}{a^2H^2}\left[ 4(8\tilde{\alpha}_M^{\rm GC}-\tilde{\alpha}_B^{\rm GC})\Phi \hat{\pi} - (\tilde{\alpha}_B^{\rm GLPV} - 16\tilde{\alpha}_M^{\rm GC} + 4\tilde{\alpha}_B^{\rm GC}) \hat{\pi}^2 + 16 \tilde{\alpha}_M^{\rm GC} \Phi^2 \right]
    \nn
    &=
    -\frac{k^2}{a^2H^2}\frac{4\mathcal{G}}{\alpha_K}
    \left[ (\alpha_B+\alpha_T-\alpha_H)\hat{\pi}+(\alpha_T-\alpha_H) \Phi \right]^2
    \nn
    &\simeq
    -4\alpha_V
    \left[ (\alpha_B+\alpha_T-\alpha_H)\hat{\pi}+(\alpha_T-\alpha_H) \Phi \right]^2
    \,.
\end{align}
Furthermore, the coefficients~$\tilde{c}_i(t,k)$ can be approximated by $c_i(t)$ where $c_i(t)$ are defined by \eqref{def_c1}--\eqref{def_c6} by replacing $\tilde{\alpha}_i(t,k)$ with $\alpha_i(t)$. Hence, the function~$\alpha_V$ changes the coefficients of $\Phi\hat{\pi}$, $\hat{\pi}^2$, and $\Phi^2$. The quadratic action formally takes the same form as that of the scalar-tensor EFT and the field equations are derived by following the calculations of the scalar-tensor theories~\cite{Gleyzes:2014qga,Kobayashi:2014ida,DAmico:2016ntq,Hirano:2018uar,Crisostomi:2017pjs,Hirano:2019nkz}. However, the explicit results are lengthy due to the presence of the coupling~$\dot{\Phi}\hat{\pi}$ and it might be difficult to extract general features of the vector-tensor EFT. Therefore, we shall focus on special cases here and leave the general analysis for future studies.

Assuming $\alpha_H=0$ and $\alpha_B,\alpha_K,\alpha_T,\alpha_V=\mathcal{O}(1)$, the coupling~$\dot{\Phi}\hat{\pi}$ is absent in the sub-horizon limit and thus the field equations of the gravitational potentials and $\hat{\pi}$ are easily derived. In particular, the gravitational potentials under the quasi-static approximation are determined by
\begin{align}
    \frac{k^2}{a^2} \Psi &= -\mu (t)4\pi G \bar{\rho}_{\rm m} \Delta 
    \label{Poisson_eq} \,, \\
    \eta (t) & = \frac{\Phi}{\Psi}
    \label{slip_para}
    \,.
\end{align}
with $8\pi G=1/M^2(t_0)$ where $t_0$ is a reference time. The effective gravitational coupling and the slip parameter are given by
\begin{align}
    \mu(t) &=\frac{M^2(t_0)}{M^2(t)}\left[ 1+ \alpha_T -2\alpha_T^2 \alpha_V + \frac{2}{V_S}( \alpha_M + \mathcal{A}-2\mathcal{A}\alpha_T \alpha_V)^2 \right]
    \,, \label{mu:H=0} \\
    \eta(t) &=\frac{1}{\mu(t)} \frac{M^2(t_0)}{M^2(t)}\left[ 1+ \frac{2(\mathcal{A}+\alpha_T)}{V_S (1+\alpha_T)}
    ( \alpha_M + \mathcal{A}-2\mathcal{A}\alpha_T \alpha_V) \right]
    \,, \label{eta:H=0}
\end{align}
where $V_S$ and $\mathcal{A}$ are defined by \eqref{def_VS} and \eqref{def_calA}, respectively. Since the stability conditions require $M^2>0,~1+\alpha_T>0$, and $V_S>0$, the effective gravitational coupling is always attractive in the scalar-tensor EFT~$(\alpha_V=0)$. On the other hand, the function $\alpha_V$ is positive and enters $\mu(t)$ with the negative sign in the third term of \eqref{mu:H=0}, implying that the vector-tensor EFT typically weakens gravity in comparison with its scalar-tensor counterpart. A realization of weak gravity was pointed out by \cite{DeFelice:2016uil} in the generalized Proca theory and we reproduce the same conclusion based on our EFT framework.

However, the expression \eqref{mu:H=0} concludes $\mu(t_0)>1$ if we impose the constraint on the speed of gravitational waves, $c_T^2\equiv 1+\alpha_T=1$~(i.e.~$\bar{M}_3^2=0$). We thus consider another case~$\alpha_V \gg 1$ without assuming $\alpha_H=0$ to find a characteristic modification of the vector-tensor EFT. In this case, we cannot ignore the $k$-dependence of $\mathcal{G}$ since the scale~$\gf^2M_2^4=\alpha_V \alpha_K H^2$ is deep inside the horizon. Nonetheless, under $\alpha_V \gg 1$, the first line of \eqref{LDE_SS} is always dominant regardless of whether $k^2/(a^2H^2) \gg \alpha_V \alpha_K$ or $1 \ll k^2/(a^2H^2) \ll \alpha_V \alpha_K$. Then, the approximate solution of $\hat{\pi}$ is found to be
\begin{align}
    \hat{\pi} = -\frac{\alpha_T-\alpha_H}{\alpha_B+\alpha_T-\alpha_H} \Phi + \mathcal{O}(\alpha_V^{-1})
    \,.
    \label{pi_sol}
\end{align}
We then substitute the solution~\eqref{pi_sol} to the Lagrangian~\eqref{LDE_SS}. Interestingly, the resultant quadratic action is independent of $\mathcal{G}$ at the leading order. The gravitational potentials are determined by the same equations as \eqref{Poisson_eq} and \eqref{slip_para} with
\begin{align}
        \eta^{-1}(t) &= 1 +  \frac{\alpha_H-\alpha_T}{\mathcal{A}+\alpha_B \alpha_T}\Biggl\{ \alpha_M - \frac{\alpha_B}{\mathcal{A}}\left(\alpha_M + \alpha_T \frac{\dot{H}}{H^2} \right)
        \nn
        & \hspace{3.5cm}
        +\frac{\alpha_H-\alpha_T}{\mathcal{A}} \left[ \frac{\bar{\rho}_{\rm m}}{2M^2H^2} + \frac{\dot{H}}{H^2}
        - \frac{1}{H}\frac{\D}{\bar{N}\D t}\left( \frac{\alpha_B \alpha_T}{\alpha_H-\alpha_T} \right) \right] \Biggl\}
        \,,  \label{eta:alphaV} 
        \\
        \mu(t)  &= \frac{1}{\eta(t)} \frac{M^2(t_0)}{M^2(t)} \left( 1 + \frac{\alpha_B \alpha_T}{\mathcal{A}} \right)
        \,,
        \label{mu:alphaV}
\end{align}
under the limit~$\alpha_V \gg 1$. Note that the effective gravitational coupling and the slip parameter are scale-independent under the quasi-static approximation despite the existence of the additional scale~$\gf^2 M_2^4~( \gg H^2)$. Taking the limit $\alpha_V \gg 1$, the results~\eqref{mu:H=0} and \eqref{eta:H=0} agree with \eqref{mu:alphaV} and \eqref{eta:alphaV} for $\alpha_H=0$. 

Let us discuss general properties of modification of gravity in the vector-tensor EFT under $\alpha_V \gg 1$. First of all, the slip parameter is modified only if $\alpha_H-\alpha_T \propto \mu_1^2 \neq 0$. Hence, observational constraints on $\eta$ can read constraints on the presence of the operator~$\delta \tilde{g}^{00} \spatialR$ in the vector-tensor EFT with $\alpha_V \gg 1$. On the other hand, the effective gravitational coupling $\mu$ is modified either when $\alpha_H-\alpha_T \propto \mu_1^2 \neq 0$ or $\alpha_B \alpha_T \neq 0$. In particular, imposing $c_T^2\equiv 1+\alpha_T=1$, the effective gravitational coupling and the slip parameter satisfy the simple relation
\begin{align}
        \mu(t) \eta(t) = 
        \frac{f(t_0)}{f(t)}
        \,,
        \label{mueta_cT=1}
    \end{align}
    in which $\mu(t)$ is simplified to be
    \begin{align}
    \mu(t) &= \frac{f(t_0)}{f(t)}\left[ 1 + \frac{\alpha_H}{(\alpha_B-\alpha_H)^2}\left\{ \alpha_H \left(\frac{\bar{\rho}_{\rm m}}{2M^2H^2}+\frac{\dot{H}}{H^2}\right) + \alpha_M (\alpha_H- 2\alpha_B) \right\} \right]
    \,. \label{mu:T=0}
\end{align}
The vector-tensor EFT may either strengthen or weaken gravity since the stability conditions do not constrain the sign of the second term of \eqref{mu:T=0}.

\section{Linear vector perturbations}
\label{sec:vector_pert}

We consider vector perturbations around the background configuration defined by the FLRW metric~\eqref{FLRW-BG} and the time-dependent temporal component for the background vector field~$\bar{A}_{\mu} = \bar{A}_0(t) \delta^0{}_\mu$. The vector perturbations are given by
\begin{equation}
    \delta g_{0i}=\bar{N} a \beta_i \,, \qquad
    \delta{g}_{ij}=a^2 ( \partial_{i}C_{j} + \partial_{j}C_{i} ) \,, \qquad \delta A_\mu=\left( 0, A_i \right) \,,
\end{equation}
with the constraints~$\delta^{ij}\partial_i\beta_j = \delta^{ij}\partial_iC_j = \delta^{ij}\partial_i A_j = 0$. The freedom of the spatial diffs allows us to set $C_i=0$ as a gauge choice, which we adopt throughout the following calculations. Since the vector perturbations cannot form a scalar quantity at the linear order, we find
\begin{align}
    \delta_1 \tilde{g}^{00}=\delta_1 \tilde{K}=\delta_1 \! \spatialR =0
    \,,
\end{align}
for the vector perturbations. Therefore, the relevant action for the linear vector perturbations is given by
\begin{align}
    S=\int \D^4x \sqrt{-g} \Bigg[
    &\frac{\Mpl^2}{2}f(t) \tilde{R} - \Lambda(t)-d(t)\tilde{K} +
    \nn
    &
    - \frac{1}{2} {\bar M}_3^2(t) \delta \tilde{K}^{\alpha}{}_{\beta} \delta \tilde{K}^{\beta}{}_{\alpha} 
- \frac{1}{4} \gamma_1(t) F_{\alpha\beta} F^{\alpha\beta}  
- \frac{1}{4} \gamma_2(t) {\tilde F}_{\alpha\beta} {\tilde F}^{\alpha\beta} 
+ \cdots \Bigg] + S_{\rm m}\,,
\end{align}
where $\Lambda(t)$ and $d(t)$ are determined by the background equations of motion.

As for the matter sector, we assume the perfect fluid. For the vector perturbations, the dynamics of the (gauge-invariant) velocity perturbations of the perfect fluid is determined by the conservation law, independently from the metric perturbations. The dynamics of the rotational component of perfect fluid is modified only through the modification of the background dynamics and the velocity perturbations will decay in time at the late time. Therefore, we shall ignore the vector-type perturbations of the matter field and use the k-essence as the matter field, similarly to the analysis in subsection~\ref{sec:stability}.

The shift perturbations~$\beta_i$ are non-dynamical. After integrating out $\beta_i$, the quadratic action for the vector perturbations is given by
\begin{align}
    \delta_2 S=\int \frac{\D t \D^3k}{(2\pi)^3} \bar{N}a \frac{\gamma_1}{2}
    \left[ \dot{A}_i^2
    -\left( c_V^2 \frac{k^2}{a^2} + m_V^2 \right)A_i^2 \right]
    \,,
\end{align}
with
\begin{align}
    c_V^2&=1+\frac{\gamma_2}{\gamma_1}+\frac{1}{2}\alpha_V \alpha_T (1+\alpha_T)
    \,, \\
    m_V^2&= \alpha_V H^2\left[ \frac{\bar{\rho}_{\rm m} + \bar{p}_{\rm m}}{M^2H^2}+2(1+\alpha_T)\left(\frac{\dot{H}}{H^2}-\alpha_M \right)-\frac{2\dot{\alpha}_T}{H} \right]
    \,,
\end{align}
where $M^2(t)$, $\alpha_T(t)$, $\alpha_M(t)$, and $\alpha_V(t)$ are defined by \eqref{def_M}, \eqref{def_alpha}, \eqref{def_alphaM}, and \eqref{def_alphaV}, respectively. The ghost and gradient instabilities are absent for
\begin{align}
    \gamma_1>0 \,, \qquad c_V^2>0
    \,.
\end{align}
The mass of the vector perturbations can be written as
\begin{align}
    m_V^2=-\frac{\alpha_V}{M^2}\left(\dot{d}+4\Mpl^2 \dot{f} H \right) \,,
\end{align}
by the use of the background EFT coefficients, meaning that the sign of $m_V^2$ is determined by the background (recall that the stability conditions impose $\alpha_V>0$ and $M^2>0$). In particular, the theories with $\dot{d}=\dot{f}=0$ yield the $\Lambda$CDM background in which the vector perturbation is massless. Note that we do not need to exclude the presence of tachyonic instability, $m_V^2<0$, because the instability exists only in low-momentum modes just like the Jeans instability. Rather, depending on the time-dependence of $\gamma_1$, the negative value of $m_V^2$ would lead to a collapse of dark energy on large scales which could yield some observational signatures of vector-tensor EFT of dark energy.

\section{Concluding remarks}\label{summary}

We have investigated the effective field theory (EFT) of vector-tensor theories. The spacetime diffeomorphism invariance is assumed to be spontaneously broken by the existence of a preferred timelike direction, in which the residual symmetries are the spatial diffeomorphism invariance and the combined $U(1)$ and time diffeomorphism invariance \eqref{U(1)time_intro}. The structure of the EFT action~\eqref{EFT_L} has been determined by the symmetry breaking pattern, the perturbative expansion around the cosmological background, and the derivative expansion. In comparison with the conventional EFT of inflation/dark energy, the residual symmetry is enlarged thanks to the gauge field. The additional symmetry gives rise to a series of consistency relations between the otherwise arbitrary EFT coefficients. In the weak coupling limit of the gauge field~($\g \to 0$), the additional symmetry recasts as the global shift symmetry of the scalar field (see Figure~\ref{fig:webofeft}). We have clarified that three classes of models, i.e.~the general scalar-tensor theories, the shift-symmetric scalar-tensor theories, and the vector-tensor theories, are distinguished by the consistency relations and the phenomenological function~$\alpha_V(t)$. As an application of our EFT, we have established a unified description of the linear cosmological perturbations. The previously developed techniques for the scalar-tensor theories can be extended to the vector-tensor theories by replacing the time-dependent functions according to \eqref{def_Meff2}--\eqref{def_lambda1} or \eqref{alpha_relation}. The phenomenological function~$\alpha_V$ typically weakens gravity (in accordance with the result of \cite{DeFelice:2016uil}) and determines the mass of the vector perturbations.

There are various directions for future studies. The biggest advantage of our EFT formulation is that one can treat the vector-tensor theories on an equal footing with the scalar-tensor theories with a clear boundary. One may then systematically obtain observational constraints on both scalar-tensor and vector-tensor theories in a single framework. From the theoretical point of view, one may clarify the relations between our EFT formulation and the proposed vector-tensor theories and may generalize the EFT to include (to be more precise, stop ignoring) additional operators, say operators relevant for generic degenerate theories called extended vector-tensor theories~\cite{Kimura:2016rzw}. It would be also interesting to understand the connection between the EFT and underlying UV physics, e.g.~following the idea of~\cite{Cheng:2006us,Mukohyama:2006mm}. In the scalar-tensor case, recent studies discuss the consistency conditions with its underlying UV physics in the broken spacetime symmetry from bottom-up~\cite{Baumann:2015nta,Grall:2021xxm} and top-down~\cite{Aoki:2021ffc} perspectives along the line of positivity bounds~\cite{Adams:2006sv}.\footnote{Currently, the bottom-up arguments and the top-down arguments suggest different constraints on EFT without the Lorentz symmetry. See~\cite{Aoki:2021ffc} for more discussions.} These consistency conditions can put additional constraints/boundaries on the EFT coefficients. In the era of precision cosmology, we may test underlying assumptions of dark energy/modified gravity models by understanding the boundaries of the theory space of the EFT.

\vspace{0.7cm}

{\bf Acknowledgments:} We would like to thank Francesco Di Filippo for initial collaboration and useful discussions. The work of K.A.~was supported by JSPS KAKENHI Grant No.~19J00895 and No.~20K14468. The work of M.A.G.~was supported by Japan Society for the Promotion of Science~(JSPS) Grants-in-Aid for international research fellow No.~19F19313. The work of S.M.~was supported in part by JSPS Grants-in-Aid for Scientific Research No.~17H02890, No.~17H06359, and by World Premier International Research Center Initiative, MEXT, Japan. The work of K.T.~was supported by JSPS KAKENHI Grant No.~JP21J00695.
\vspace{0.7cm}

\appendix

\section{\texorpdfstring{The $1 + 3$ decomposition with vorticity}{The 1 + 3 decomposition with vorticity}}
\label{app:1+3}
In this appendix, we summarize the relations between four-dimensional tensors and their parallel and orthogonal objects with respect to a unit timelike vector~$\tilde{n}^{\mu}$ in the presence of vorticity. Although there is no global hypersurface orthogonal to $\tilde{n}^{\mu}$, we can define several geometrical objects in the tangent hyperplane at each point. The projection tensor is defined as
\begin{align}
    \tilde{h}_{\mu\nu} \equiv g_{\mu\nu}+\tilde{n}_{\mu} \tilde{n}_{\nu}
    \,.
\end{align}
The derivative of $\tilde{n}^{\mu}$ is decomposed into
\begin{align}
    \nabla_{\mu} \tilde{n}_{\nu} = \tilde{B}_{\mu\nu} - \tilde{n}_{\mu} \tilde{a}_{\nu}
    \,,
\end{align}
with
\begin{align}
    \tilde{B}_{\mu\nu} & \equiv \tilde{h}^{\alpha}{}_{\mu} \nabla_{\alpha} \tilde{n}_{\nu} \,,
    \\
    \tilde{a}_{\mu} &\equiv \tilde{n}^{\alpha} \nabla_{\alpha}\tilde{n}_{\mu}
    \,.
\end{align}
The tensor~$\tilde{B}_{\mu\nu}$ is further decomposed into the kinematical quantities, namely the expansion, the shear, and the vorticity, as
\begin{align}
    \tilde{B}_{\mu\nu}=\tilde{K}_{\mu\nu} + \tilde{\omega}_{\mu\nu} 
    =\frac{1}{3}\tilde{K}\tilde{h}_{\mu\nu} + \tilde{\sigma}_{\mu\nu} + \tilde{\omega}_{\mu\nu} 
    \,,
\end{align}
where
\begin{align}
    \tilde{K}_{\mu\nu} &\equiv \tilde{B}_{(\mu\nu)}\,, \\
    \tilde{\omega}_{\mu\nu} & \equiv \tilde{B}_{[\mu\nu]} \,, \\
    \tilde{K} &\equiv \tilde{K}^{\mu}{}_{\mu} =\tilde{B}^{\mu}{}_{\mu}\,, \\
    \tilde{\sigma}_{\mu\nu} &\equiv \tilde{K}_{\mu\nu}-\frac{1}{3}\tilde{K} \tilde{h}_{\mu\nu}
\,.
\end{align}

Let $\tilde{T}^{\mu \cdots}{}_{\nu \cdots}$ be a projected object of a four-dimensional tensor~$T^{\alpha\cdots}{}_{\beta \cdots}$, that is, 
\begin{align}
\tilde{T}^{\mu \cdots}{}_{\nu \cdots} \equiv \tilde{h}^{\mu}{}_{\alpha}\cdots \tilde{h}^{\beta}{}_{\nu}\cdots T^{\alpha \cdots}{}_{\beta \cdots}
\,.
\end{align}
We then define the orthogonal spatial derivative~$\tilde{D}_{\mu}$ as
\begin{align}
    \tilde{D}_{\mu} \tilde{T}^{\nu \cdots}{}_{\rho \cdots} 
    \equiv \tilde{h}^{\alpha}{}_{\mu}\tilde{h}^{\nu}{}_{\beta}\cdots \tilde{h}^{\gamma}{}_{\rho} \cdots
    \nabla_{\alpha} \tilde{T}^{\beta \cdots}{}_{\gamma \cdots} 
    \,.
\end{align}
The orthogonal spatial derivative can be represented by
\begin{align}
    \tilde{D}_{\mu} \tilde{T}^{\nu \cdots}{}_{\rho \cdots} 
    = \tilde{\partial}_{\mu} \tilde{T}^{\nu \cdots}{}_{\rho \cdots}
    +\tilde{\Gamma}^{\nu}_{\alpha \mu} \tilde{T}^{\alpha \cdots}{}_{\rho \cdots} + \cdots
    - \tilde{\Gamma}^{\alpha}_{\rho \mu} \tilde{T}^{\nu \cdots}{}_{\alpha \cdots} - \cdots
    \,,
\end{align}
where $\tilde{\partial}_{\mu}$ and $\tilde{\Gamma}^{\mu}_{\nu\rho}$ are the orthogonal spatial partial derivative and the projected connection defined by
\begin{align}
    \tilde{\partial}_{\mu} \tilde{T}^{\nu \cdots}{}_{\rho \cdots} 
    &\equiv \tilde{h}^{\alpha}{}_{\mu}\tilde{h}^{\nu}{}_{\beta}\cdots \tilde{h}^{\gamma}{}_{\rho} \cdots
    \partial_{\alpha} \tilde{T}^{\beta \cdots}{}_{\gamma \cdots} 
    \,,
    \\
    \tilde{\Gamma}^{\mu}_{\nu\rho} & \equiv \tilde{h}^{\mu}{}_{\alpha}\tilde{h}^{\beta}{}_{\nu} \tilde{h}^{\gamma}{}_{\rho} \Gamma^{\alpha}_{\beta \gamma}
    \,.
    \label{project_Gamma}
\end{align}
Here, $\Gamma^{\alpha}_{\beta\gamma}$ is the four-dimensional Levi-Civita connection. 
Note that the orthogonal spatial partial derivative is defined not only for a projected object of some four-dimensional tensor but also for any three-dimensional object, e.g.~$\tilde{\Gamma}^{\mu}_{\nu\rho}$.
One can easily confirm that the orthogonal spatial covariant derivative is compatible with the projection tensor,
\begin{align}
    \tilde{D}_{\mu} \tilde{h}_{\nu\rho} = 0
    \,.
\end{align}

The objects analogous to the curvature and the torsion associated with $\tilde{D}_{\mu}$ may be defined as~\cite{Roy:2014lda} 
\begin{align}
    {}^{(3)}\! \tilde{R}^{\mu}{}_{\nu\rho\sigma} & \equiv
    2\left( \tilde{\partial}_{[\rho|} \tilde{\Gamma}^{\mu}_{\nu |\sigma]} 
    + \tilde{\Gamma}^{\mu}_{\lambda [\rho|} \tilde{\Gamma}^{\lambda}_{\nu | \sigma]} \right)  + \tilde{h}^{\mu}{}_{\beta} \tilde{h}^{\gamma}{}_{\nu} \tilde{\rm T}^{\alpha}{}_{\rho\sigma} \Gamma^{\beta}_{ \gamma \alpha}
    -2\tilde{h}^\mu{}_\alpha\tilde{h}^\beta{}_\nu\tilde{h}^\gamma{}_{[\rho}\tilde{h}^\delta{}_{\sigma]}(\partial_\gamma\tilde{n}^\alpha)\partial_\delta\tilde{n}_\beta
    \,, \label{orthogonal_spatial_curvature} \\
    \tilde{\rm T}^{\alpha}{}_{\mu\nu} &\equiv - 2\tilde{n}^{\alpha} \tilde{\omega}_{\mu\nu}
    \,,
\end{align}
by which the commutators of the orthogonal spatial derivatives are written as
\begin{align}
    2 \tilde{D}_{[\mu} \tilde{D}_{\nu]} f &= - \tilde{\rm T}^{\alpha}{}_{\mu\nu} \nabla_{\alpha}f
    \,, \\
    2 \tilde{D}_{[\mu} \tilde{D}_{\nu]} \tilde{V}^{\rho}
    & = {}^{(3)}\! \tilde{R}^{\rho}{}_{\sigma \mu \nu} \tilde{V}^{\sigma}
    - \tilde{h}^{\rho}{}_{\beta} \tilde{\rm T}^{\alpha}{}_{\mu\nu} \nabla_{\alpha} \tilde{V}^{\beta}
    \,, \label{commutator_vector}
\end{align}
for a scalar~$f$ and a projected vector~$\tilde{V}^{\mu}$, respectively. 
The relation~\eqref{project_Gamma} immediately concludes $\tilde{\Gamma}^{\mu}_{[\nu\rho]}=0$ while the non-commutativity of $\tilde{\partial}_{\mu}$, namely $\tilde{\partial}_{[\mu} \tilde{\partial}_{\nu]} \neq 0$, leads to the non-vanishing commutation relation even for a scalar.
The commutators are reminiscent of those of the Riemann-Cartan space. However, we emphasize that the superscript of the ``torsion'' tensor~$\tilde{\rm T}^{\alpha}{}_{\mu\nu}$ is parallel to the vector~$\tilde{n}^{\alpha}$, meaning that $\tilde{\rm T}^{\alpha}{}_{\mu\nu}$ is not a three-dimensional object. Since the derivative operator~$\tilde{D}_{\mu}$ is defined in the tangent hyperplane at each point and not on a three-manifold, we need a reference to the four-dimensional spacetime to define the objects~${}^{(3)}\! \tilde{R}^{\mu}{}_{\nu\rho\sigma}$ and $\tilde{\rm T}^{\alpha}{}_{\mu\nu} $. 

We shall refer to ${}^{(3)}\! \tilde{R}^{\mu}{}_{\nu\rho\sigma}$ as the orthogonal spatial curvature. The orthogonal spatial Ricci tensor and scalar are defined by
\begin{align}
\spatialR_{\mu\nu} \equiv \spatialR^{\alpha}{}_{\mu\alpha\nu}\,, \quad 
\spatialR \equiv \spatialR^{\mu}{}_{\mu}
\,,
\end{align}
where we note that $\spatialR_{\mu\nu}$ is not symmetric in its indices.
The orthogonal spatial curvature and the four-dimensional curvature is related via the Gauss equation,
\begin{align}
    \tilde{h}^{\alpha}{}_{\mu} \tilde{h}^{\beta}{}_{\nu} \tilde{h}^{\gamma}{}_{\rho} \tilde{h}^{\delta}{}_{\sigma} R_{\alpha\beta\gamma\delta} =
    {}^{(3)}\!\tilde{R}_{\mu\nu\rho\sigma} + 2 \tilde{B}_{[\rho|\mu} \tilde{B}_{|\sigma]\nu}
    \,.
\end{align}
In addition, the Codazzi equation and the Ricci equation are
\begin{align}
\tilde{h}^{\alpha}{}_{\mu} \tilde{n}^{\beta} \tilde{h}^{\gamma}{}_{\rho} \tilde{h}^{\delta}{}_{\sigma} R_{\alpha\beta\gamma \delta } &=
2\tilde{D}_{[\rho}\tilde{B}_{\sigma]\mu}-2\tilde{a}_{\mu}\tilde{\omega}_{\rho\sigma}
\,,
\label{Codazzi_eq} \\
\tilde{h}^{\alpha}{}_{\mu} \tilde{n}^{\beta} \tilde{h}^{\gamma}{}_{\rho} \tilde{n}^{\delta} R_{\alpha\beta\gamma \delta } &= - \pounds_{\tilde{n}} \tilde{B}_{\rho\mu} + \tilde{B}_{\mu \lambda} \tilde{B}_{\rho}{}^{\lambda} + \tilde{D}_{\rho}\tilde{a}_{\mu} + \tilde{a}_{\mu}\tilde{a}_{\rho}
\,,
\label{Ricci_eq}
\end{align}
respectively. Therefore, all the components of the four-dimensional curvature are written in terms of the kinematical quantities and the orthogonal spatial curvature. In particular, the four-dimensional Ricci scalar and the Raychaudhuri equation are
\begin{align}
    R&={}^{(3)}\! \tilde{R} + 2\pounds_{\tilde{n}}\tilde{K}+\tilde{B}_{\mu\nu}\tilde{B}^{\nu\mu}+\tilde{K}^2 -2 \tilde{D}_{\mu}\tilde{a}^{\mu}
    -2\tilde{a}_{\mu}\tilde{a}^{\mu}
    \nn
    &={}^{(3)}\! \tilde{R} + \tilde{K}_{\mu\nu}\tilde{K}^{\mu\nu}-\tilde{K}^2 -\tilde{\omega}_{\mu\nu}\tilde{\omega}^{\mu\nu} -2 \nabla_{\mu}(\tilde{a}^{\mu} -\tilde{K}\tilde{n}^{\mu})
\,,
\end{align}
and
\begin{align}
     R^{\mu\nu}\tilde{n}_{\mu}\tilde{n}_{\nu}
     &= -\pounds_{\tilde{n}}\tilde{K}-\tilde{B}_{\mu\nu}\tilde{B}^{\nu\mu}+\tilde{D}_{\mu}\tilde{a}^{\mu}
     +\tilde{a}_{\mu}\tilde{a}^{\mu}
         \nn
     &= \tilde{K}^2-\tilde{K}_{\mu\nu}\tilde{K}^{\mu\nu}+\tilde{\omega}_{\mu\nu}\tilde{\omega}^{\mu\nu}+\nabla_{\mu}(\tilde{a}^{\mu}-\tilde{K}\tilde{n}^{\mu})
     \,,
\end{align}
respectively.

Let us show that the orthogonal spatial curvature is decomposed into three irreducible pieces.
The indices of the orthogonal spatial curvature satisfy the same symmetry as the curvature in the Riemann-Cartan space:
\begin{align}
    \spatialR_{[\mu\nu]\rho\sigma}=\spatialR_{\mu\nu[\rho\sigma]}=\spatialR_{\mu\nu\rho\sigma}
    \,,
\end{align}
and
\begin{align}
    \spatialR_{\mu[\nu\rho\sigma]}=-2\tilde{B}_{[\nu|\mu} \tilde{B}_{|\rho\sigma]}=-2\tilde{B}_{[\nu|\mu} \tilde{\omega}_{|\rho\sigma]}\neq 0
    \,.
    \label{antisymmetricR}
\end{align}
Thus, the orthogonal spatial curvature is decomposed into three independent blocks without using the metric, which is represented by
\begin{align}
    \ydiagram{1,1} \otimes \ydiagram{1,1} &= \ydiagram{2,2} \oplus \ydiagram{2,1,1} \oplus \ydiagram{1,1,1,1} 
    \label{decomposition}
\end{align}
by means of the Young diagrams. The Young diagram of shape $(2,2)$ [the first term in the right-hand side of \eqref{decomposition}] is the piece that has the same symmetry group as the Riemann curvature, so it can be further decomposed into three irreducible pieces by the use of the metric, namely the Ricci scalar, the symmetric trace-free part of the Ricci tensor, and the Weyl tensor. Recall that $\spatialR_{\mu\nu\rho\sigma}$ is a three-dimensional object in the sense that $\spatialR_{\mu\nu\rho\sigma}$ is orthogonal to $\tilde{n}^{\alpha}$. Since the Weyl piece vanishes identically in three dimensions, we only have two irreducible pieces from the first term in the right-hand side of \eqref{decomposition}, which are
\begin{align}
    \spatialR \qquad {\rm and} \qquad
    \spatialR^{T}_{\mu\nu} \equiv \spatialR_{(\mu\nu)}-\frac{1}{3}\tilde{h}_{\mu\nu}\spatialR
    \,.
\end{align}
In addition, tensors with more than three antisymmetric indices vanish in three dimensions, meaning that
\begin{align}
    \ydiagram{2,1,1} \oplus \ydiagram{1,1,1,1} = \ydiagram{1}\,,
\end{align}
where we have used
\begin{align}
    \ydiagram{1,1,1}=1\,, \quad  \ydiagram{1,1,1,1}=0 \,.
\end{align}
Hence, the second and third terms in the right-hand side of \eqref{decomposition} give one irreducible piece which is given by \eqref{antisymmetricR} or
\begin{align}
    \spatialR^V_{\mu}\equiv \spatialR_{\mu\nu\rho\sigma}\tilde{\epsilon}^{\nu\rho\sigma} \,,
\end{align}
in the vector representation, where $\tilde{\epsilon}^{\nu\rho\sigma}\equiv \tilde{n}_{\mu}\epsilon^{\mu\nu\rho\sigma}$ and $\epsilon^{\mu\nu\rho\sigma}$ is the four-dimensional Levi-Civita tensor. In summary, all the components of the orthogonal spatial curvature are represented by three irreducible pieces,
\begin{align}
    \spatialR\,,\quad \spatialR^{T}_{\mu\nu}\,, \quad  \spatialR^V_{\mu}
    \,.
\end{align}
The irreducible decomposition of the orthogonal spatial curvature is explicitly given by
\begin{align}
    \spatialR_{\mu\nu\rho\sigma}= \frac{1}{3}\spatialR\, \tilde{h}_{\mu[\rho}\tilde{h}_{\sigma] \nu} 
    + 2\left( \spatialR^{T}_{\mu[\rho}\tilde{h}_{\sigma]\nu}-\spatialR^{T}_{\nu[\rho}\tilde{h}_{\sigma]\mu} \right) 
    + \frac{1}{2}\spatialR^V_{[\mu} \tilde{\epsilon}_{\nu]\rho\sigma}
    \,.
\end{align}

When the vorticity vanishes, the Frobenius theorem concludes that the vector~$\tilde{n}_{\mu}$ is hypersurface orthogonal. We then denote the unit vector orthogonal to hypersurfaces by $n_{\mu}$. The quantities~$\tilde{D}_{\mu}$ and ${}^{(3)}\!\tilde{R}_{\mu\nu\rho\sigma}$ are reduced to the spatial covariant derivative~$D_{\mu}$ and the spatial curvature~${}^{(3)}\! R_{\mu\nu\rho\sigma}$ associated with the spatial metric~$h_{\mu\nu}\equiv g_{\mu\nu}+n_{\mu}n_{\nu}$ on the three-manifold.

\section{Operators with Lie derivative}
\label{sec:Lie}
\setcounter{equation}{0}
\renewcommand{\theequation}{B\arabic{equation}}

When the EFT Lagrangian contains the Lie derivative, the Lagrangian should be also expanded in terms of the Lie derivatives of the building blocks. We define 
\begin{align}
    \delta \pounds_{\tilde{n}} \tilde{g}^{00} \equiv \pounds_{\tilde{n}} \tilde{g}^{00} - (\pounds_{\tilde{n}}\tilde{g}^{00})_{\rm BG}(t) \,,
\end{align}
and so on, where $(\pounds_{\tilde{n}}\tilde{g}^{00})_{\rm BG}(t)$ is the background part of $\pounds_{\tilde{n}}\tilde{g}^{00}$. Terms non-linear in perturbations of the Lie derivatives give new operators in the Lagrangian~\eqref{EFT_L}, say $(\delta \pounds_{\tilde{n}} \tilde{g}^{00})^2$. In this Appendix, we particularly focus on terms which are linear in Lie derivatives.

The Taylor expansion of $\mathcal{L}_0$ may have an additional linear terms
\begin{align}
    \mathcal{L}_0 \ni \barL_{\pounds_{\tilde{n}}\tilde{g}^{00} }(t) \delta \pounds_{\tilde{n}}\tilde{g}^{00} = \barL_{\pounds_{\tilde{n}}\tilde{g}^{00} }(t)  \pounds_{\tilde{n}}\tilde{g}^{00} - \barL_{\pounds_{\tilde{n}}\tilde{g}^{00} }(t) (\pounds_{\tilde{n}}\tilde{g}^{00})_{\rm BG}(t)
    \,,
\end{align}
with the Taylor coefficient~$\barL_{\pounds_{\tilde{n}}\tilde{g}^{00} }(t)$. In general, we have the Lie derivatives of the other building blocks of \eqref{background_blocks} and higher-order Lie derivatives which we write as $\pounds_{\tilde{n}}T$ collectively where $T=\{ \tilde{g}^{00}, \tilde{K}, \cdots \}$ is an invariant scalar. Note that $\mathcal{L}_2$ does not provide the terms of the form~$f(t)\pounds_{\tilde{n}}T$ because we have eliminated \eqref{FLT} by using integration by parts. The linear term takes the form
\begin{align}
    \int \D^4x \sqrt{-g} f(t) \pounds_{\tilde{n}}T
    =-\int \D^4 x \sqrt{-g} (\tilde{n}^{\mu}\partial_{\mu}f(t) + \tilde{K}f(t) )T
    \,,
\end{align}
by performing integration by parts where $f(t)$ is the corresponding Taylor coefficient. In the scalar-tensor EFT, the quantity~$n^{\mu}\partial_{\mu}f=\sqrt{-g^{00}}\partial_t f$ can be absorbed into other terms. In the vector-tensor case, on the other hand, because of $\tilde{n}^{\mu}\partial_{\mu}f(t)\neq \sqrt{-\tilde{g}^{00}}\partial_t f$, the quantity $\tilde{n}^{\mu}\partial_{\mu}f(t)$ is not given by the building blocks, providing a new operator of the EFT action.

Let us split $T$ into the background and the perturbation parts, $T=T_{\rm BG}(t)+\delta T$. The term~$\tilde{n}^{\mu}\partial_{\mu}f(t) T_{\rm BG}(t) $ can be absorbed to the term~$\int \D^4x \sqrt{-g} d(t)\tilde{K}=-\int \D^4x \sqrt{-g}  \tilde{n}^{\mu}\partial_{\mu}d(t)$. We then consider the residual component
\begin{align}
    \int \D^4x \sqrt{-g} \tilde{n}^{\mu}\partial_{\mu}f(t) \delta T = \int \D^4x \sqrt{-g}  \left[ (\tilde{n}^{\mu}\partial_{\mu}f)_{\rm BG}(t) \delta T +  \partial_t f \delta \tilde{n}^{0} \delta T  \right]\,,
\end{align}
where $(\tilde{n}^{\mu}\partial_{\mu}f)_{\rm BG}(t)$ is the background part of $\tilde{n}^{\mu}\partial_{\mu}f(t)$ and $\delta\tilde{n}^0=\tilde{n}^0-\tilde{n}^0_{\rm BG}(t)$. The first term starts at the linear order in perturbations but it can be absorbed into the operator~$g(t)T$. When $T(t)$ is either $\tilde{g}^{00}, \tilde{K}$, or $\spatialR$, it is already present in the EFT action~\eqref{EFT_L}. On the other hand, we need to continue the same calculation if $T$ is given by the Lie derivative of a scalar. In either case, the term~$(\tilde{n}^{\mu}\partial_{\mu}f)_{\rm BG}(t) \delta T$ can be handled systematically and then the background dynamics of the EFT is determined by $f(t)$, $\Lambda(t)$, $d(t)$, and $c(t)$. On the other hand, the second term~$\delta \tilde{n}^0 \delta T$ provides a new contribution to the perturbations.

As a result, the general EFT of vector-tensor theories may have the following operators:
\begin{align}
    \delta \tilde{n}^0 \delta \tilde{g}^{00}\,, \quad \delta \tilde{n}^0 \delta \tilde{K}\,, \quad \delta \tilde{n}^0 \delta \spatialR
    \,,  \quad \cdots \,.
\end{align}
Since we have $\delta_1 \tilde{n}^0=0$ under the decoupling limit of gravity, i.e.~under the ansatz~\eqref{decoupling_metric} and \eqref{decoupling_vector}, the additional operators are irrelevant for linear perturbations if the metric perturbations are completely decoupled from the perturbations of the preferred vector. On the other hand, the coupling cannot be ignored in the context of dark energy. 
In particular, the irrotational ansatz leads to $\delta_1 \tilde{n}^0 = \delta_1 n^0 = -\frac{\bar{N}}{2}\delta_1 g^{00}$ in the gauge~$A_{\mu}=A_0 \delta^0_{\mu}$, meaning that the quadratic action additionally has
\begin{align}
    \delta_1 g^{00} \delta_1 \tilde{g}^{00}\,, \quad \delta_1 g^{00} \delta_1 K\,, \quad \delta_1 g^{00} \delta_1 \spatialRST
    \,, \quad \cdots \,,
    \label{additional_op}
\end{align}
from the Lie derivatives. Although we have assumed that the Lie derivative is suppressed by $\cutoff$ in the main text, the operators~\eqref{additional_op} are not higher-derivative ones and can be added to the EFT action even at the leading order.

\bibliographystyle{JHEPmod}
\bibliography{refs}

\end{document}